\documentclass[epj,nopacs]{svjour}

\usepackage[T1]{fontenc} 
\usepackage[dvipsnames]{xcolor}

\usepackage{graphics}
\usepackage[utf8]{inputenc}
\usepackage{xspace}
\usepackage{amsmath}
\usepackage{amssymb}
\usepackage{url}
\usepackage{rotating}
\usepackage{enumerate}
\usepackage{graphicx}
\usepackage[compat=1.1.0]{tikz-feynman}
\usepackage{pdflscape, afterpage, capt-of} 
\usepackage{slashed} 
\usepackage{multirow}
\usepackage[normalem]{ulem}
\usepackage{xfrac}
\usepackage[export]{adjustbox} 
\usepackage{makecell} 
\usepackage{subcaption} 
\usepackage{bbold}
\usepackage{stackengine} 
\usepackage{subcaption}
\usepackage{bm}

\numberwithin{equation}{subsection} 

\usepackage{def} 

\newcommand{\nc}{\ensuremath{N_{C}}}                
\newcommand{\nf}{\ensuremath{N_{F}}}                
\newcommand{\pid}{\ensuremath{\pi_D}}               
\newcommand{\rhod}{\ensuremath{\rho_D}}             
\newcommand{\mpi}{\ensuremath{m_{\pi_D}}}           
\newcommand{\mrho}{\ensuremath{m_{\rho_D}}}         
\newcommand{\lamd}{\ensuremath{\Lambda_{D}}}        
\newcommand{\zp}{\ensuremath{Z^\prime}}             
\newcommand{\Psplit}{\ensuremath{P_{\mathrm{split}}}}
\newcommand{\pT}{\ensuremath{p_{T}}}

\title{Dark Sector Showers and Hadronisation in Herwig 7}

\author{Suchita Kulkarni\inst{1}\and M.R. Masouminia\inst{2} \and Simon Plätzer\inst{1,3} \and Dominic Stafford\inst{4}
}                     
\institute{Institute of Physics, NAWI Graz, University of Graz, Universitätsplatz 5, A-8010 Graz, Austria
  \and
  IPPP, Department of Physics, University of Durham, South Road, Durham DH1 3LE, United Kingdom
  \and
        Particle Physics, Faculty of Physics, University of Vienna, Boltzmanngasse 5, A-1090 Wien, Austria
\and Deutsches Elektronen-Synchrotron DESY, Notkestr. 85, 22607 Hamburg, Germany
      }
\date{\today}

\abstract{We present a novel simulation of a strongly interacting dark
  sector also known as the Hidden Valley scenarios using angular ordered showers and the cluster hadronisation
  model in Herwig 7. We discuss the basics of this implementation and
  the scale hierarchies underpinning the simulation. With the help of
  a few benchmarks, we show the effect of variation of dark sector
  parameters on thrust and angularities within the dark sector, and
  study correlation functions, which can be helpful for understanding
  the angular structure of these events.  Finally we comment on the
  uncertainties introduced due to lack of knowledge of hadronisation
  parameters within the dark sectors.}

\begin{document} 

\maketitle


\section{Introduction}
\label{sec:introduction}

Standard Model (SM) extensions featuring new confining non-Abelian sectors~\cite{Strassler:2006im,Han:2007ae}  coupled with the SM via some portal present an exciting opportunity for new physics searches at colliders as they produce unique, previously unexplored signatures in the form of anomalous jets. The non-Abelian sectors could feature any gauge group, number of colours or flavours~\cite{Kulkarni:2022bvh,Beauchesne:2019ato,Pomper:2024otb,Hochberg:2014kqa,Kamada:2022zwb}. Moreover, they may or may not contain matter fields~\cite{Batz:2023zef,Soni:2016gzf,Curtin:2022oec,Forestell:2016qhc,Soni:2017nlm,Carenza:2022pjd}. Finally, the matter field content and their representation can also be chosen freely, e.g. scalar, Weyl or Dirac fermions in fundamental, (anti-) symmetric or adjoint representation. In this work, we concentrate on QCD-like confining Hidden Valley scenarios, where the non-Abelian sector in the ultra-violet (UV) contains Dirac fermions ($q_D$) in the fundamental representation of $SU(\nc)$ gauge group featuring a chiral symmetry breaking in the infra-red (IR).  These extensions are generically dubbed confining Hidden Valleys and the class signatures containing anomalous jets is called dark showers, akin to SM QCD showering and hadronisation process. 

For systems in the chirally broken phase, the dark shower paradigm and corresponding generation of `dark jets' can be thought of analogous to the SM jet generation, where dark quarks produced in the hard process e.g. $ p p \to q_D \bar{q}_D$ undergo rapid parton showering and subsequent hadronisation. Some of the hadronised bound states decay back to the SM via the portal producing visible signatures.

Dark showers and their associated phenomenology have gained signifiant attention in the recent years. The experimental signatures include semi-visible jets~\cite{Cohen:2015toa,Cohen:2017pzm,Beauchesne:2017yhh}, lepton-jets~\cite{ArkaniHamed:2008qp,Baumgart:2009tn,Chan:2011aa,Buschmann:2015awa}, emerging jets~\cite{Schwaller:2015gea,Renner:2018fhh} and among the extreme signatures such as soft-unclustered energy patters~\cite{Polchinski:2002jw,Hatta:2008tx,Knapen:2016hky,Harnik:2008ax}. Results from first experimental searches for semi-visible jets are also available~\cite{CMS:2021dzg,ATLAS:2023swa}. For a review on strongly-coupled theories see e.g.~\cite{Albouy:2022cin,Kribs:2016cew,Cacciapaglia:2020kgq}. 

Given the rich theoretical and phenomenological landscape presented by confining Hidden Valleys, a systematic exploration is necessary. Among the requirements, development of reliable event generators, used to analyse the experimental signatures is important.  Dark showers have so far been simulated using the {\tt PYTHIA} Hidden Valley framework~\cite{Carloni:2010tw,Carloni:2011kk,Bierlich:2022pfr}, which underwent extensive validation and improvements during the Snowmass process~\cite{Albouy:2022cin}. It is important to note that unlike the SM there is no possibility of tuning the empirical dark shower hadronisation parameters to data. This presents a source of uncertainty in dark shower predictions, and moving away from unconstrained phenomenological parameters ultimately calls for a re-thinking of hadronisation models \cite{Platzer:2022jny,Hoang:2024zwl}. At this point, different models are therefore desirable for comparison, and we present an extension of the Herwig~7 event generator for simulating new strongly coupled sectors resulting in dark showers at colliders. In addition we present first phenomenological results using the framework implemented in Herwig, and study some observables which might serve as novel experimental constraints and as theoretical benchmarks to monitor the accuracy of the dark shower simulation algorithms and hadronisation models. Similar investigations for QCD hadronisation models would be beneficial and could create more confidence in simulations of dynamics in the dark sectors.

This paper is organised as follows: in section~\ref{sec:herwig} we present the setup of the simulation of dark sectors in the Herwig~7 event generator, with associated shower and hadronisation modules. We in particular point out the underlying assumptions regarding hierarchies of energy scales. In section~\ref{sec:benchmark} we present results for a benchmark scenario, and in section~\ref{sec:phenomenology} we carry out some initial phenomenological investigations, including a large class of promising observables such as angularities and correlation functions. Finally in section~\ref{sec:had_param_var} we investigate the impact of variations of the parton shower and hadronisation tuning parameters on the predictions.

\section{Dark Sector in Herwig}
\label{sec:herwig}

\subsection{Herwig Angular Ordered Shower with Dark Sector Branchings}
\label{sec:showers}

Herwig~7 \cite{Bahr:2008pv,Bellm:2015jjp,Bewick:2023tfi} is a multi-purpose event generator, which features two parton shower algorithms, an angular ordered one based on QCD coherence \cite{Gieseke:2003rz}, and a dipole shower \cite{Platzer:2009jq,Platzer:2011bc}. In particular the former has recently been extended to include EW interactions in addition to QCD \cite{Masouminia:2021kne,Darvishi:2021het,Lee:2023hef}, possibly interleaved with QCD radiation. In this case we have used the new shower interactions to simulate radiation in a dark sector, which is otherwise decoupled from the Standard Model parton evolution.

The angular ordered shower is accurate for all those processes in which observables are azimuthally integrated around a hard jet axis, and for which coherence can be exploited by commensurate angular hierarchies among the jets, or due to being completely inclusive about the jet structure. Further improvements for subsequent large-angle soft radiation can be implemented by suitably including azimuthal correlations in the evolution \cite{Richardson:2018pvo}. The Herwig implementation of the angular ordered shower should therefore provide an accurate picture of the dark radiation pattern. A comparison with other algorithms is desirable, but certainly beyond the scope of the present study. Since we start from a hard process involving a hard quark-antiquark pair charged under the dark sector, no non-trivial colour correlations, which would need to be handled in terms of the large-$N$ limit, appear. Additionally, the initial conditions are set such that there is a symmetric partition of the radiation phase space in between the two showers developing off the dark quarks from the decay of the mediator. Mass effects in the parton showers in Herwig are generally understood and predicted very well \cite{Cormier:2018tog} and have recently been understood also at the analytic level \cite{Hoang:2018zrp} including accuracy considerations for the angular ordered shower in general \cite{Bewick:2019rbu,Bewick:2021nhc}.

\subsection{The Cluster Hadronisation Model in Light of Additional Strong Interactions}
\label{sec:hadronisation}

Hadronisation of the dark showers has been accomplished by a generalisation of the Herwig cluster hadronisation model \cite{Webber:1983if} as described in \cite{Herwigpp}. To achieve this, the hadronisation handlers were generalised, removing the explicit SM dependence, such as separate parameters for c and b quarks, and replacing this with possibility to set these parameters for an arbitrary spectrum of heavy quarks: the ``HadronSelector'' class was generalised to a ``HadronSpectrum'' class containing any remaining model specific information. One can therefore create a set of handlers dedicated to hadronisation of the particles charged under SM QCD, and a second set for hadronisation of the dark sector.  The ``DarkHadronSpectrum'' can handle up to 9 dark quarks, which can be either light (i.e. have mass significantly smaller than the dark confinement scale $\lamd$, similar to the SM u, d and s quarks) or heavy (similar to the SM c and b quarks). So far only production of dark mesons has been studied, and the production of dark Baryons will become available in a future release.

Being semi-empirical, the hadronisation model has a number of parameters, which for SM QCD can be tuned to obtain the best possible agreement with data. For the dark sector this is not possible, so these parameters must be set to estimates of reasonable values, based on their physical meaning and the values found to give a good fit for QCD. The most relevant of these parameters will be discussed here, as well as some recommendations for well-motivated values.

The first set of relevant parameters are the so-called constituent masses of the dark quarks and gluon, which control the mass these particles can be considered to have in the hadronisation (which will in general differ from the current quark masses). For dark quarks this should be greater than half the mass of the lightest meson which they can form to ensure their clusters have sufficient energy to decay to into at least one dark hadron, while for the dark gluons it must be greater than twice the constituent mass of the lightest dark quark to allow these to be split by the ``PartonSplitter'' class. At the start of the hadronisation all dark gluons are split into dark quark - anti-quark ($q_{D}^{i}\bar{q}_{D}^{i}$) pairs. The rate at which these split into a particular flavour $i$ is controlled by a parameter $Pwt_{\mathrm{Split}}^{i}$. This should be set to one for the lightest dark quarks, and zero for significantly heavier dark quarks that are very unlikely to be pair-produced; it can also take a value in-between for a dark quark with a relatively small mass splitting from the lightest dark quarks, similarly to the strange quark in the SM, though this scenario has not yet been explored.

At this point the colour connections of the quarks can be reconnected to reduce the mass of the clusters, including the formation of baryonic clusters. For SM QCD this has been demonstrated to improve the predictions of the fraction of baryons produced \cite{Gieseke:2017clv}, however since the production of dark baryons has not yet been investigated this setting should be left off for now.

Clusters which are too heavy will fission into lighter clusters. This threshold is controlled by two parameters, and fission occurs if a cluster exceeds a mass of
\begin{equation}
   M =  \left(Cl_{\mathrm{max}}^{Cl_{\mathrm{pow}}}+ \left(m_{1}+m_{2}\right)^{Cl_{\mathrm{pow}}}\right)^{1/Cl_{\mathrm{pow}}},
\end{equation}
where $m_{1}$ and $m_{2}$ are the constituent masses of the quarks in the cluster, and $Cl_{\mathrm{max}}$ and $Cl_{\mathrm{pow}}$ are parameters of the model, which are the same for clusters containing only the lightest dark quarks, but can be set separately for clusters containing heavier dark quarks. $Cl_{\mathrm{max}}$ represents the highest mass for which a cluster can reasonably be considered a pseudo-hadronic bound state; while the exact value this would take for the spectrum of dark hadrons is unclear, in the SM this ranges from about $17\Lambda_{QCD}$ for clusters containing only light quarks to about $20\Lambda_{QCD}$ for clusters containing b quarks, so a similar multiple of $\lamd$ would make sense for the dark sector. The best value for $Cl_{\mathrm{pow}}$ is unclear; for the SM a value of 2.78 is found to work well for light clusters, but the best fit value for bottom clusters is 0.547. However if all dark quarks are mass degenerate varying $Cl_{\mathrm{pow}}$ will be equivalent to varying $Cl_{\mathrm{max}}$, so one can leave $Cl_{\mathrm{pow}}$ at a fixed value and set the threshold based on $Cl_{\mathrm{max}}$ alone.

Clusters fission through creating a $q_{D}^{i}\bar{q}_{D}^{i}$ pair from the vacuum, with the flavour determined by a set of parameters $Pwt_{\mathrm{Fission}}^{i}$ similar to $Pwt_{\mathrm{Split}}^{i}$. Since the energy scale for decay is lower than for the splitting of the gluons from the parton shower, strange-like dark quarks will be slightly more suppressed compared to the lightest dark quarks, so a lower value of $Pwt_{\mathrm{Fission}}^{i}$ compared to $Pwt_{\mathrm{Split}}^{i}$ would be appropriate. The masses of the two outgoing clusters, $M_{1}$ and $M_{2}$ are then given by power-like distributions,
\begin{equation}
  \frac{{\rm d}P}{{\rm d}M_i} \sim \left(M_i - m \right)^{\Psplit - 1} \ ,
\end{equation}
including the kinematic constraints $M_1+M_2 < M$, and $M_i > m_i + m$, where $m_i$ are the constituent masses of the original cluster, and $m$ the one belonging to the created quark pair.

\Psplit{} is a tuning parameter. The best value of \Psplit, which will affect how much of the parent cluster masses is converted to mass of the child clusters, compared to their momentum, is unclear, and varies in the SM between 0.625 (for bottom clusters) and 0.994 (for charm clusters). However the SM value for light clusters, 0.899, seems a good starting point - this will lead to most of the energy being converted to mass of the child clusters.

The remaining important set of parameters which need to be set control the decay of clusters to two hadrons. Similarly to cluster fissioning, this involves producing a $q_{D}^{i}\bar{q}_{D}^{i}$ pair with flavour determined by parameters $Pwt^{i}$. The same general principles apply when selecting the values of $Pwt^{i}$ as $Pwt_{\mathrm{Split}}^{i}$ and $Pwt_{\mathrm{Fission}}^{i}$, however the energy scale for decay is again lower, so dark quarks heavier than the lightest ones will likely be further suppressed. There is also an option to produce two $q_{D}^{i}\bar{q}_{D}^{i}$ pairs, which in an $SU(3)$ theory will lead to the production of baryons rather than mesons, however since we have not yet explored dark baryons we recommend leaving the parameter controlling this, $Pwt^{\mathrm{Diquark}}$, set to 0 for now.

Generally, the hadrons produced in cluster decays are emitted isotropically, however it is found for the SM a better description of the kinematics is achieved if the hadrons which contain a quark from the perturbative parts of the calculation (i.e. not from gluon or heavy cluster splittings) are emitted in broadly the same direction as the perturbative quark \cite{Herwigpp}. There is therefore an option to emit these hadrons in this direction, smeared by an angle $\theta$ distributed according to
\begin{equation}
  \frac{{\rm d}P}{{\rm d}\cos \theta} \sim \exp\left(\frac{\cos\theta - 1}{Cl_{\mathrm{Smr}}}\right)
\end{equation}
where $Cl_{\mathrm{Smr}}$ is a parameter which controls the degree to which the hadron direction is smeared compared to the parent quark. This option is used in the SM, with $Cl_{\mathrm{Smr}}$ varying between 0.78 for light quarks and 0.078 for bottom quarks; in the dark sector it would similarly make sense to use a reasonably large value for dark quarks with masses significantly below $\lamd$, where the direction is likely to be smeared by hadronisation effects, and small values for heavier dark quarks which are more likely to retain their direction.

\section{A Benchmark Model}
\label{sec:benchmark}

\subsection{Physical Scale Hierarchies and Parameters for the Simulation}
\label{sec:physicalscenarios}

For validation and initial phenomenological studies we followed the benchmarks outlined in the 2021 Snowmass report on dark showers \cite{Albouy:2022cin}. This family of benchmarks contain a 1 TeV spin-1 $\zp$ mediator, which couples to the SM via mixing with the photon, and $\nf$ mass-degenerate dark quarks, charged under the $U(1)_D$ group associated with the $\zp$, and under an $SU(\nc)$ group associated with the dark gluon. This leads to a relatively simple hadron spectrum, with $\nf^{2}-1$ mass-degenerate pseudo-scalar dark pions, a significantly heavier flavour singlet pseudo-scalar $\eta_{D}^{\prime}$,  $\nf^{2}-1$ mass-degenerate vector mesons, $\rhod$, and the flavour singlet vector $\omega_{D}$. The mass spectrum of all of these hadrons has been computed as a function of $\lamd$ and the mass of the dark quarks, $m_{q_{D}}$, though from a practical point of view it is easier to parameterise everything in terms of the dark pion mass, $\mpi$. Fits to non-perturbative calculations of dark hadron masses \cite{Albouy:2022cin,Fischer:2006ub} give:

\begin{eqnarray}
    m_{q_{D}} &=& 0.033 \frac{\mpi^2}{\Lambda_{D}} \nonumber \\
    m_{\omega_{D}} = \mrho &=& \sqrt{5.76 \Lambda_{D}^{2} + 1.5 \mpi^{2}} \ ,
    \label{eq:snowmass_fits}
\end{eqnarray}
while an analysis of the chiral Lagrangian \cite{Albouy:2022cin} gives
\begin{eqnarray}
    m_{\eta_{D}^{\prime}} &= \sqrt{\mpi^2+9\frac{\nf}{\nc}\Lambda_{D}^{2}} \ .
    \label{eq:eta_fit}
\end{eqnarray}

\begin{table}
\begin{center}
\begin{tabular}{ |c |c |c |c |c |} 
\hline
 & $\frac{\mpi}{\lamd}$ & $\nf$ & $\rhod$ decays & $\pid$ decays\\
\hline
Scenario A & 0.6 & 3 & $\rhod^{\rm{all}}\to\pid \pid$ & $\pid^{\text{diag}}\to f\overline{f}$\\
\hline
\multirow{2}{*}{Scenario B}  & \multirow{2}{*}{1.7} & \multirow{2}{*}{4} & $\rhod^{\text{diag}}\to f\bar{f}$ & \multirow{2}{*}{$\pid$ stable} \\
& & & $\rhod^{\text{off--diag}} \to \pid f \bar{f}$ & \\
\hline
\end{tabular}

\end{center}
\caption{Benchmarks considered in this work for numerical results. In each of the scenarios,  $\nc = 3$ and three values of $\lamd = 5,10,40\,\rm{GeV}$ were considered. The $\pid, \rhod$ masses were fixed using \eqref{eq:snowmass_fits}-\eqref{eq:eta_fit}.}
\label{tab:benchmarks}
\end{table}

There are then two broad classes of benchmarks: if $\mrho > 2\mpi$, the $\rhod$ mesons will decay to a pair of dark pions, some of which are unstable and decay to SM quarks (we will refer to this as scenario A), whereas for $\mrho < 2\mpi$ the $\rhod$ mesons will instead decay to SM quarks since the decay to dark pions is not possible, while all of the dark pions are assumed to be stable (we will refer to this as scenario B). The exact values of free parameters for both these scenarios are summarised in table~\ref{tab:benchmarks}. These benchmarks are the same as those proposed in the snowmass dark showers studies~\cite{Albouy:2022cin}, except that we only consider the case in which all diagonal dark pions (for scenario A) and rho mesons (for scenario B) promptly decay to the SM final states, and we consider slightly different values of $\lamd$. 

We do not focus on variations in the number of decaying hadrons since this is a sub-dominant effect compared to variations in \lamd\, or between scenarios, however we briefly discuss the effect of varying the number of dark mesons decaying to SM at the end of section \ref{sec:phenomenology}. The different values of \lamd\, are chosen for a variety of reasons; the $\lamd=5$ GeV benchmark features $\pid$ mesons with a mass of 3 GeV, which in this model should decay to a charm quark-antiquark pair (see section \ref{sec:decays}), however Herwig cannot handle this decay below the $D\bar{D}$ meson threshold (since the module which handles the decay requires that QCD cluster formed can decay to a pair of mesons containing the quarks) - in this case one would need to implement decay directly to the QCD mesons which would be formed by the decay of the $c\bar{c}$ pair, so we instead consider a benchmark of $\lamd=4$ GeV, where the dark hadron will decay to an $s\bar{s}$ pair. Conversely a value of $\lamd=50$ GeV is very high for the 1 TeV mediator we consider here, so we consider instead a value of $\lamd=30$ GeV. Additionally we consider a $\lamd=1$ GeV benchmark for scenario B to also investigate the impact of variations of \lamd{} in this case, though we did not consider any higher values of \lamd{} since these would have very high \mrho{} and hence few \rhod{} mesons, which are the ones which decay visibly, would be produced by a shower starting from a 1 TeV mediator.

It is possible for the decaying dark hadrons to have a long lifetime, which gives rise to emerging jet signatures. This functionality is implemented in Herwig, however we did not investigate it in this work since these emerging jet signatures require different analysis strategies to the prompt signatures we consider.

To take full advantage of the cluster hadronisation model, we assigned individual particle IDs to each dark hadron, rather than using combined particle IDs for all flavour diagonal and non-diagonal states, as has been done in some previous approaches. The $\eta_{D}^{\prime}$ mass was assigned using the fit to the chiral Lagrangian given in the Snowmass dark showers report \cite{Albouy:2022cin}, rather than being taken degenerate to the dark pions, and the $\eta_{D}^{\prime}$ and $\omega_{D}$ were allowed to decay to three dark pions, where this was kinematically allowed. A script implementing the full decay tables for these benchmarks is available from the authors upon request.

\subsection{Hard Process}
\label{sec:hardprocess}

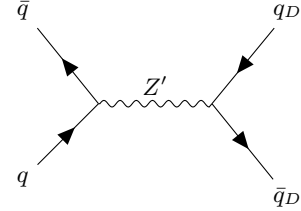
\begin{figure}[ht]
    \centering
    \begin{tikzpicture}
        \begin{feynman}
        \vertex (i1) {$q$};
        \vertex[above=1.0cm of i1] (i);
        \vertex[above=1.0cm of i] (i2) {$\bar{q}$};
        \vertex[right=1.0cm of i] (a);
        \vertex[right=1.5cm of a] (b);
        \vertex[right=1.0cm of b] (f);
        \vertex[above=1.0cm of f] (f1) {$q_{D}$};
        \vertex[below=1.0cm of f] (f2) {$\bar{q}_{D}$};
        
        \diagram*{
            (i1) -- [fermion] (a) -- [fermion] (i2),
            (a) -- [boson, edge label=\(Z^{\prime }\)] (b),
            (f1) -- [fermion] (b) -- [fermion] (f2)
        };
        \end{feynman}
    \end{tikzpicture}
    \caption{Production of a dark quark-antiquark pair via a $\zp$ mediator.}
    \label{fig:hard_process}
\end{figure}

The dark quarks were produced via s-channel production of the $\zp$ mediator, as shown in figure \ref{fig:hard_process} using the Herwig ResonantProcessConstructor. More complicated processes, such as associated production with SM particles or NLO diagrams can be simulated using an external ME provider via the Herwig Matchbox module \cite{Matchbox,Herwig7}. So far the Herwig dark shower is only compatible with s-channel models; we plan to include support for t-channel models (where the mediator is charged under both QCD and the dark strong interaction) in future iterations. 

\subsection{Parton Shower and Hadronisation parameters}
\label{sec:psparams}

\begin{figure}[!ht]
    \centering
  \begin{minipage}[b]{0.5\textwidth}
    \centering
    \includegraphics[width=8cm]{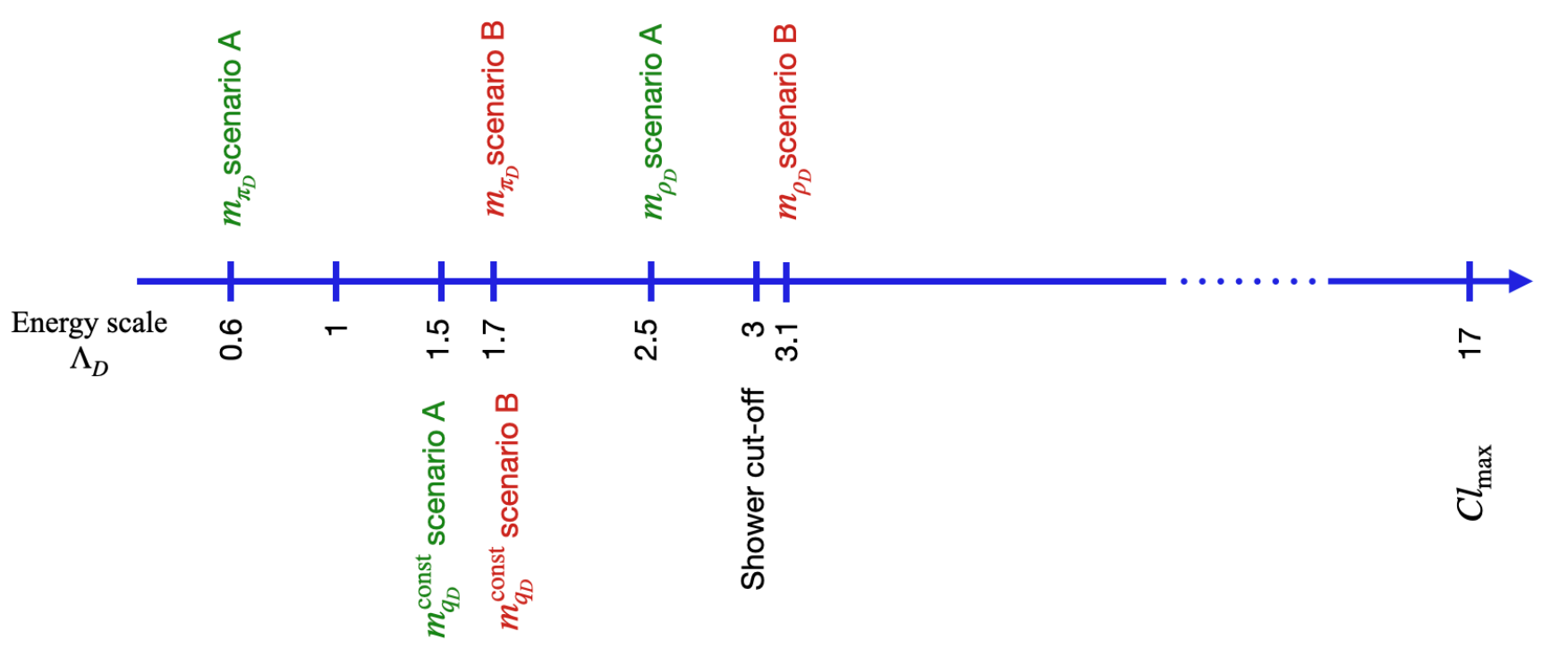}
\end{minipage}\hfill
    \caption{A schematic representation of various scales involved in the process and their relationship to the scale of the dark sector $\lamd$.}
    \label{fig:scale_hierarchies}
\end{figure}

In general the parton shower and hadronisation parameters were set to be either the same as in the Standard Model, or the same as a multiple of \lamd/$\Lambda_{\mathrm{QCD}}$, for the parameters with units of energy. For the shower this meant setting the shower \pT{} cut-off to $3\Lambda_{D}$. Herwig also allows non-perturbative evolution of $\alpha_{\mathrm{D}}$ below some scale $\mathrm{Q}_{\mathrm{min}}$, however we set this equal to the shower cut-off, meaning $\alpha_{\mathrm{D}}$ was treated perturbatively everywhere.

For the hadronisation, we generally followed the recommendations outlined in section \ref{sec:hadronisation}, with the values shown in table \ref{tab:ps_had_params}; the various $Pwt^{i}$ parameters were all set to 1.0 since the dark quarks are mass degenerate.  The least obvious parameters to set were the constituent masses; having investigated the cluster mass distributions for a number of different choices we decided to set the dark quark constituent masses for scenario A to 1.5\lamd, which is the value used in the SM, which also has $\mpi\approx0.6\Lambda_{\mathrm{QCD}}$. For scenario B one would expect the constituent masses to be slightly heavier since the hadrons are heavier; we therefore decided to set the dark quark constituent masses equal to the dark pion mass, $\mpi=1.7\lamd$, which ensures that every cluster can decay to two dark mesons. In both cases the dark gluon constituent masses were set to 2.2 times the dark quark constituent masses, which ensures the dark gluons can be split to a dark $q_{D}^{i}\bar{q}_{D}^{i}$ pair. We schematically represent these scale hierarchies in figure~\ref{fig:scale_hierarchies}. The impact of varying the most sensitive of these parameters is shown in section \ref{sec:had_param_var}.

\begin{table*}[t]
\begin{center}
{\footnotesize
\begin{tabular}{|c|c|c|}
\hline
Parameter & Herwig setting & Value \\
\hline
shower \pT{} cut-off & \verb|/Herwig/Shower/DarkPTCutOff:pTmin| & 3\lamd \\
\hline
$\mathrm{Q}_{\mathrm{min}}$ & \verb|/Herwig/Shower/AlphaDARK:Qmin| & 3\lamd \\
\hline
$q_{D}^{i}$ constituent mass & \verb|/Herwig/Particles/DarkQuarki/ConstituentMass| & 1.5\lamd/1.7\lamd \\
\hline
$g_{D}$ constituent mass & \verb|/Herwig/Particles/DarkGluon/ConstituentMass| & 3.3\lamd/3.74\lamd \\
\hline
$Pwt_{\mathrm{Split}}^{i}$ & \verb|/Herwig/Hadronization/DarkPartonSplitter:SplitPwt| & 1.0 \\
\hline
Colour Reconnection & \verb|/Herwig/Hadronization/DarkColourReconnector:ColourReconnection| & No \\
\hline
$Cl_{\mathrm{max}}$ & \verb|/Herwig/Hadronization/DarkClusterFissioner:ClMaxLight| & 17\lamd \\
\hline
$Cl_{\mathrm{pow}}$ & \verb|/Herwig/Hadronization/DarkClusterFissioner:ClPowLight| & 2.780 \\
\hline
$Pwt_{\mathrm{Fission}}^{i}$ & \verb|/Herwig/Hadronization/DarkClusterFissioner:FissionPwt| & 1.0 \\
\hline
\Psplit & \verb|/Herwig/Hadronization/DarkClusterFissioner:PSplitLight| & 0.899 \\
\hline
$Pwt^{i}$ & \verb|/Herwig/Hadronization/DarkHadSpec:Pwt| & 1.0 \\
\hline
$Pwt^{\mathrm{Diquark}}$ & \verb|/Herwig/Hadronization/DarkHadSpec:PwtDIquark| & 0.0 \\
\hline
$Cl_{\mathrm{Dir}}$ & \verb|/Herwig/Hadronization/DarkClusterDecayer:ClDirLight| & 1 \\
\hline
$Cl_{\mathrm{Smr}}$ & \verb|/Herwig/Hadronization/DarkClusterDecayer:ClSmrLight| & 0.78 \\
\hline
\end{tabular}
}
\end{center}
\caption{Parton shower and Hadronisation parameters used for the benchmark model. Since all dark quarks are mass degenerate the same parameters are used for all dark quarks $i$.}
\label{tab:ps_had_params}
\end{table*}

\subsection{Decays}
\label{sec:decays}

\begin{figure}
    \centering
  \begin{minipage}[b]{.24\textwidth}
    \centering
    \begin{tikzpicture}
        \begin{feynman}
        \vertex (i) {$\pi_{D}^{\text{diag}}$};
        \vertex[right=1.25cm of i] (a);
        \vertex[right=0.9cm of a] (b);
        \vertex[right=1.0cm of b] (f);
        \vertex[above=0.55cm of f] (f1) {$\bar{q}$};
        \vertex[below=0.55cm of f] (f2) {$q$};
        
        \diagram*{
            (i) -- [scalar] (a) -- [boson, insertion=0, edge label'=\(Z^{\prime *}\)] (b),
            (f1) -- [fermion] (b) -- [fermion] (f2)
        };
        \end{feynman}
    \end{tikzpicture}
\end{minipage}\hfill
  \begin{minipage}[b]{.24\textwidth}
    \centering
    \begin{tikzpicture}
        \begin{feynman}
        \vertex (i) {$\rho_{D}^{\text{diag}}$};
        \vertex[right=1.25cm of i] (a);
        \vertex[right=0.9cm of a] (b);
        \vertex[right=1.0cm of b] (f);
        \vertex[above=0.55cm of f] (f1) {$\bar{q}$};
        \vertex[below=0.55cm of f] (f2) {$q$};
        
        \diagram*{
            (i) -- [boson] (a) -- [boson, insertion=0, edge label'=\(Z^{\prime *}\)] (b),
            (f1) -- [fermion] (b) -- [fermion] (f2)
        };
        \end{feynman}
    \end{tikzpicture}
\end{minipage}\hfill \\
\begin{minipage}[b]{.35\textwidth}
    \centering
    \begin{tikzpicture}
        \begin{feynman}
        \vertex (i) at (-2.5,0) {$\rho_{D}^{\text{off--diag}}$};
        \vertex (a) at (-0.8,0);
        \vertex (b) at (0.6,-0.75);
        \vertex (f1) at (1.6,-0.375) {$\bar{q}$};
        \vertex (f2) at (1.6,-1.125) {$q$};
        \vertex (f3) at (1.8, 1.0) {$\pi_{D}^{\text{off--diag}}$};
        
        \diagram*{
            (i) -- [boson] (a) -- [boson, edge label'=\(Z^{\prime *}\)] (b),
            (a) -- [scalar] (f3),
            (f1) -- [fermion] (b) -- [fermion] (f2)
        };
        \end{feynman}
    \end{tikzpicture}
\end{minipage}
    \caption{Possible decays of dark mesons to SM quarks: decays of a flavour diagonal $\pid^{0}$ (left) and $\rhod^{0}$ (centre) mesons to an SM $q\bar{q}$ pair, and, right, decay of a flavour off-diagonal $\rhod^{\pm}$ meson to a lighter dark pion and a SM $q\bar{q}$ pair.}
    \label{fig:dark_meson_decays}
\end{figure}
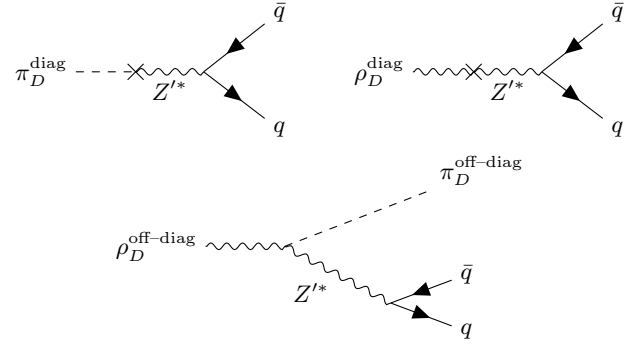

There are two possible ways in which dark hadrons can decay: if a decay to other dark hadrons is kinematically possible, this will typically be favoured over decays to SM particles, which are suppressed since they occur via mixing with the heavy $\zp$ mediator. These decays can be implemented in Herwig using the same decay matrix elements as used for the corresponding decays in the SM, for instance the Vector2Meson decayer for $\rhod\to\pid\pid$, which contains the full ME for this process. If the dark hadrons are unstable but no decay within the dark sector is kinematically allowed, they can decay to SM particles by mixing with the (off-shell) $\zp$ mediator. Since most dark shower models have dark hadrons significantly higher than the QCD confinement scale, these will not decay directly to SM hadrons, but instead to SM quarks, which will undergo further parton showering before hadronising again. The flavour diagonal $\pid^{0}$ and $\rhod^{0}$ mesons (including the $\omega_{D}$ meson) can, depending on the dark quark charges,  mix with the mediator, decaying to a SM $q\bar{q}$ pair, while off-diagonal $\rhod^{\pm}$ mesons may decay to an off-diagonal $\pid^{\pm}$ and a SM $q\bar{q}$ pair, as shown in figure \ref{fig:dark_meson_decays}.

A new DarkoniumDecayer class, based on the existing QuarkoniumDecayer, was implemented to handle these decays, including ensuring the correct colour connections. This does not rely on the details of the decay mediator, and can handle any decays of dark hadrons to a SM $q\bar{q}$ pair or a $q\bar{q}$ pair plus another dark hadron. So far the DarkoniumDecayer only implements flat phase space decays, however it is planned to include non-trivial phase space dependence in future iterations.

\section{First Phenomenological Studies}
\label{sec:phenomenology}
To investigate the structure of dark shower events in a clean environment we considered production of a pair of dark quarks via $\zp$ mediator at a 1 TeV electron-positron collider. Events with the $\zp$ mediator pair producing one of the five lightest SM quark flavours were used as a background of QCD jets at a similar energy (we found relatively little difference in the distributions considered between the quark flavours). For these studies we considered jets clustered using the anti-kt algorithm with a radius of 0.8, requiring $p_{T}>10$ GeV. All plots were produced using custom analyses implemented in the Rivet framework \cite{Rivet}.

\begin{figure}[!ht]
    \centering
    \includegraphics[width=0.45\textwidth]{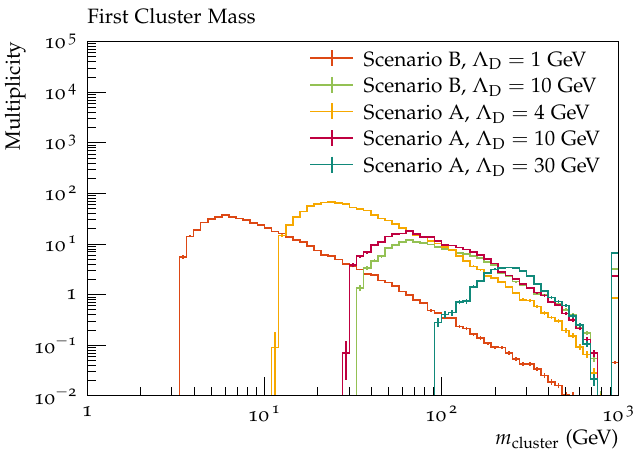}
    \includegraphics[width=0.45\textwidth]{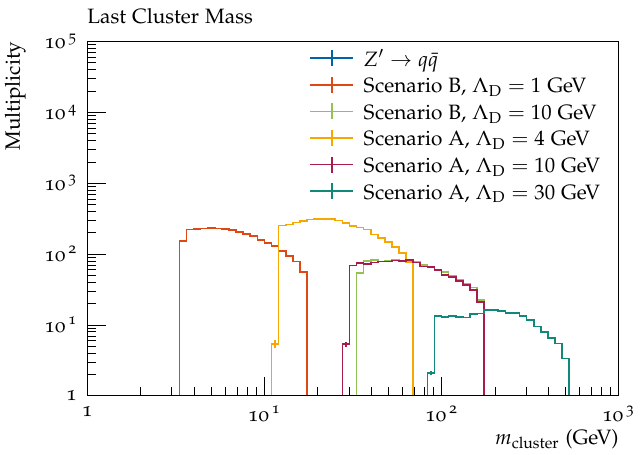}
    \caption{Cluster masses of the dark clusters in the cluster hadronisation model before cluster fissioning (top) and after (bottom).}
    \label{fig:cluster_masses}
\end{figure}

We started by investigating various internal variables in Herwig to check the models behaved as expected. One useful variable to investigate is the mass of the dark hadronic clusters, both before and after the cluster splitting. These are shown in figure \ref{fig:cluster_masses}. The lower end of cluster masses both before and after cluster fissioning is set to be twice the dark quark constituent mass. Since  the constituent dark quark mass is proportional to $\lamd$ (see table~\ref{tab:ps_had_params}), the lower end of cluster mass is in turn proportional to $\lamd$. The maximum cluster mass on the other hand differs between before fission to after fission stage. The highest cluster mass before cluster fissioning is controlled by the mass of the $\zp$ mediator (1000 GeV), which corresponds to the case there are no parton shower emissions, so the two dark quarks from the hard process form a single cluster. After the fissioning there are more clusters, within the expected range from twice the dark quark constituent mass to $Cl_{\mathrm{max}}$. Beyond the obvious differences of the different energy scales for different values of \lamd, one can also see differences between the scenario A and B benchmarks by comparing the \lamd=10 GeV benchmarks, with scenario A having lighter clusters, due to the different $\alpha_{D}$ running (principally originating from the lower \nf) and the lower constituent masses giving a lighter lower edge to the cluster mass distribution.

\begin{figure}[!ht]
    \centering
    \centering
    \includegraphics[width=0.45\textwidth]{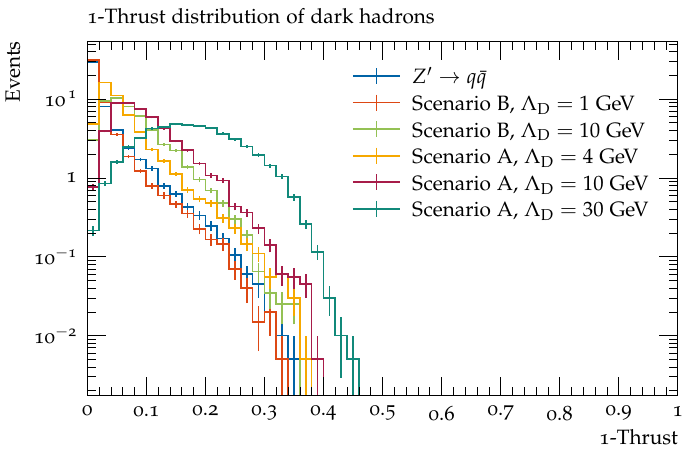}
    \caption{The 1-Thrust distribution for the dark hadrons before decays to Standard Model quarks for different dark shower benchmarks, as well as the distribution for the SM hadrons in a $\zp\to q\bar{q}$ process.}
    \label{fig:thrust_dark}
\end{figure}

Next we investigated the structure of the events in the dark sector (i.e. before the additional complication of semi-visible decays to SM particles). The variable we found most useful for this purpose was the trust, defined as:

\begin{equation}
T=\max_{\hat{n}}\frac{\sum_{i}|\bm{p}_{i}\cdot\hat{n}|}{\sum_{i}|\bm{p}_{i}|}
\end{equation}

where $\bm{p}_{i}$ is the momentum of the $i^{\mathrm{th}}$ particle, $i$ runs over all particles in the event, and $\hat{n}$ is the unit vector which maximises the thrust (the thrust axis). The $1-T$ distribution is shown in figure \ref{fig:thrust_dark}; one can see that the $\lamd=1$ GeV benchmark is very similar to the SM distribution (i.e., generally a dijet topology), while benchmarks with higher \lamd{} tend to have higher values of $1-T$, indicating a more isotropic distribution of particles. This is due to two mechanisms: firstly the parton shower emissions will be harder for higher \lamd{}, and secondly for the heavier benchmarks there will be a higher fraction of very heavy clusters formed of quarks in different hemispheres, which will fission multiple times, producing hadrons isotropically between the original quarks (see also some similar observations for the current Herwig model in \cite{Hoang:2018zrp}). Comparing the 10 GeV benchmarks, scenario A is slightly more isotropically distributed than scenario B , which can be attributed to two factors: firstly the hadron masses in scenario A are lower, and hence more of the energy of the decaying clusters is converted to kinetic energy of the hadrons, increasing their angular separation, and secondly in this scenario any \rhod\, mesons will decay to \pid\, mesons which are again somewhat separated.

A further factor which impacts the final state topology is the fraction of dark hadrons which decay to visible particles. For scenario A this is simply determined by the fraction of dark pions which are allowed to decay, in this case 2 out of 8, so 25\%. For scenario B the visible decay fraction is mainly determined by the relative production of \rhod{} to \pid{} mesons; the cluster model predicts this fraction, since clusters decay to all kinematically allowed pairs of mesons, with probability equal to the number of spin states. Since the shower and hadronisation parameters are set relative to \lamd, the rate of production of \rhod{} mesons is very similar for the two scenario B benchmarks: 54.7\% for \lamd=1 GeV and 55.7\% for \lamd=10 GeV (the difference is due to slight differences in the cluster mass spectrum due to different amount of phase space for the shower and cluster fissioning relative to the starting scale of 1 TeV). However it should be noted that this fraction has a significant dependence on the shower and hadronisation parameters used: the impact of varying these is discussed in section \ref{sec:had_param_var}. The final visible decay fraction will be further influenced by the momentum of the \rhod{} mesons relative to the \pid{} mesons, and the fact that  the flavour off-diagonal \rhod{} mesons do not decay completely visibly, but instead to a \pid{} meson plus a SM $q\bar{q}$ pair. After these decays the final visible decay fraction is 35.5\% for \lamd=1 GeV and 34.6\% for \lamd=10 GeV. The visible decay fraction is expected to increase slightly when considering non-trivial phase space dependence in the off-diagonal \rhod{} decays, since the SM $q\bar{q}$ pair will tend to carry more of the momentum from the decay. 

\begin{figure}[!ht]
    \centering
    \centering
    \includegraphics[width=6cm]{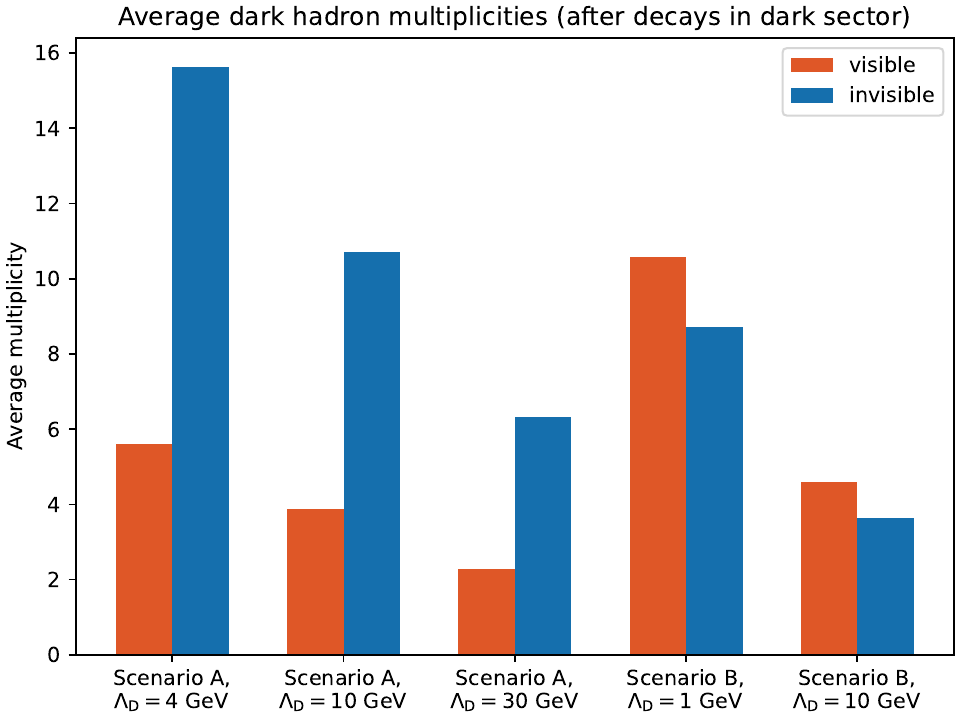}
    \caption{The average multiplicity of dark hadrons for the different benchmark scenarios, after decays within the dark sector (i.e., $\rhod\to\pid\pid$ for scenario A), divided into those which decay at least partially to SM particles (flavour-diagonal \pid{} for scenario A, and all \rhod{} for scenario B), and those which are invisible (stable).}
    \label{fig:NDarkHad}
\end{figure}

\begin{figure}[!ht]
    \centering
  \begin{minipage}[b]{.29\textwidth}
    \centering
    \includegraphics[height=2.8cm]{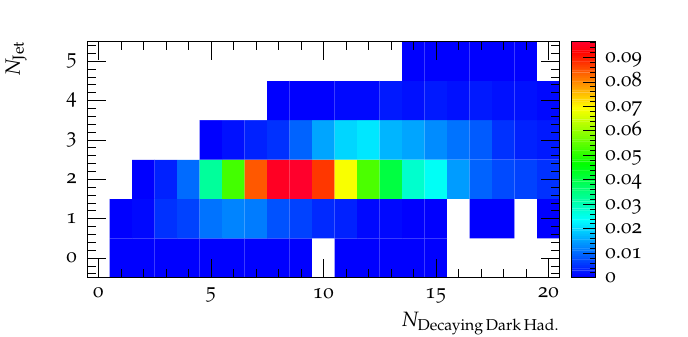}
\end{minipage}\hfill
\begin{minipage}[b]{.19\textwidth}
    \centering
    \includegraphics[height=2.8cm]{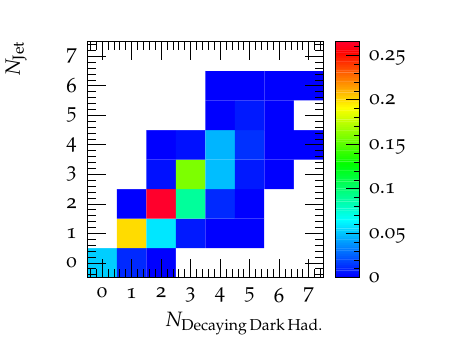}
\end{minipage}
    \caption{The distribution of events as a function of number of jets and number of hadrons decaying to visible particles for the $\lamd=1$ GeV, $\mrho < 2\mpi$ benchmark (left) and for the $\lamd=30$ GeV, $\mrho >2 \mpi$ benchmark (right).}
    \label{fig:NDarkDecNJets}
\end{figure}

While the visible decay fraction is very similar within the two scenarios, the number of dark hadrons varies significantly as a function of \lamd, as shown in figure \ref{fig:NDarkHad}. A result of the different scale is that similar signatures can appear as a result of different physics considerations; figure \ref{fig:NDarkDecNJets} shows the number of visible jets  against the number of dark hadrons which decay to SM particles (either all $\rho_{D}$ mesons for scenario B, or the flavour diagonal $\pi_{D}$ mesons for scenario B). For the $\lamd=1$ GeV benchmark there are a relatively high number of dark hadrons decaying to SM quarks, however these tend to only form two visible jets, whereas for the $\lamd=30$ GeV benchmark there are far fewer dark hadrons which decay, and each of these tends to give rise to its own visible jet (and in a few cases more than one since the decay products do not always fall within the 0.8 jet radius), however the most common case is again to observe two jets since this is the most probable number of visibly decaying dark hadrons.

\begin{figure}[!ht]
    \centering
    \centering
    \includegraphics[width=0.45\textwidth]{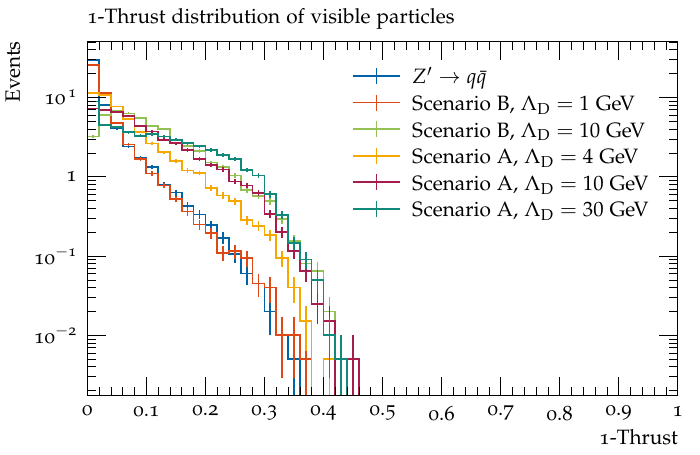}
    \caption{The 1-Thrust distributions of the visible decay products of the dark shower benchmarks and the $\zp\to q\bar{q}$ process.}
    \label{fig:thrust_vis}
\end{figure}

Looking at the thrust distributions of the visible particles (figure \ref{fig:thrust_vis}), the intermediate mass benchmarks seem to have a longer tail of high $1-T$ events compared to the distributions of all dark hadrons, since the smaller number of visible decay products can obscure the overall two jet structure. Conversely, the peak of the distribution for the \lamd=30 GeV shifts to lower values, as these events often have two visible jets since typically only two dark hadrons decay visibly. 
 
To differentiate between jets originating from the decays of dark hadrons and the SM QCD jets, it can be useful to study variables sensitive to angular structures. One new class of variables which can be particularly instructive are correlation functions, where the angular distance between each pair of visible final state particles, $\theta_{ij}$ is plotted, weighted by the energies of the two particles, $E_{i}E_{j}$. A related type of correlation functions, which sum over all angles weighted by the energies, has previously been studied in the context of dark sectors simulated with Pythia8 HV module~\cite{Cohen:2020afv}. Taking the sum of these angles gives a single variable, which can be useful when designing analyses, but can obscure substructure at different scales, hence we prefer to just plot all of the angles separately.

\begin{figure}[!ht]
    \centering
    \includegraphics[width=0.45\textwidth]{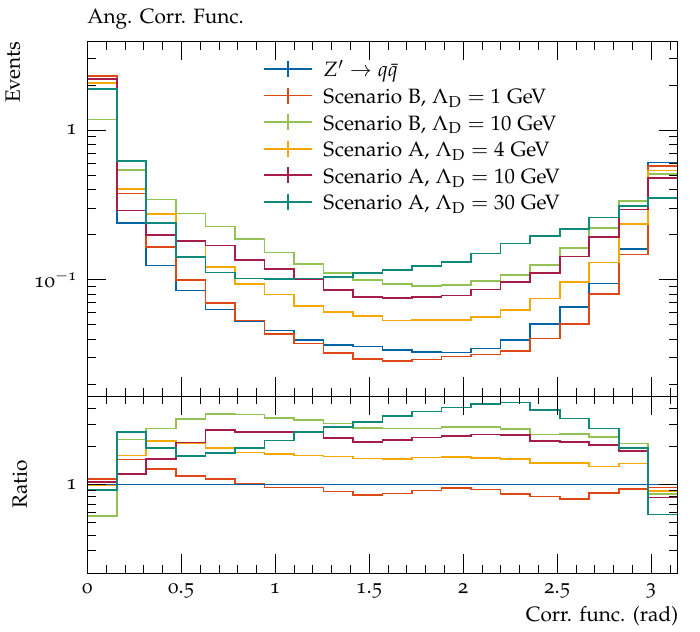}
    \caption{The correlation functions for all visible particles in events for the considered dark shower benchmarks and pair production of SM quarks by the $\zp$ mediator. }
    \label{fig:corr_across_event}
\end{figure}

By plotting the correlation functions across the event (figure \ref{fig:corr_across_event}) one can see that SM QCD and the lower $\lamd$ benchmarks are sharply peaked at 0 and $\pi$, as one would expect for a two jet topology, whereas the higher $\lamd$ benchmarks have a significantly higher amount of intermediate angular correlations, reflecting the fact that the dark hadrons are more evenly spread around by the dark hadronisation. For the $\lamd=30$ GeV benchmark there is an additional peak relative to the SM distribution at an angular scale of about 0.25, which likely corresponds to the angular separation between the quark anti-quark pairs produced from the decays of dark pions.

\begin{figure}[!ht]
    \centering
    \includegraphics[width=0.45\textwidth]{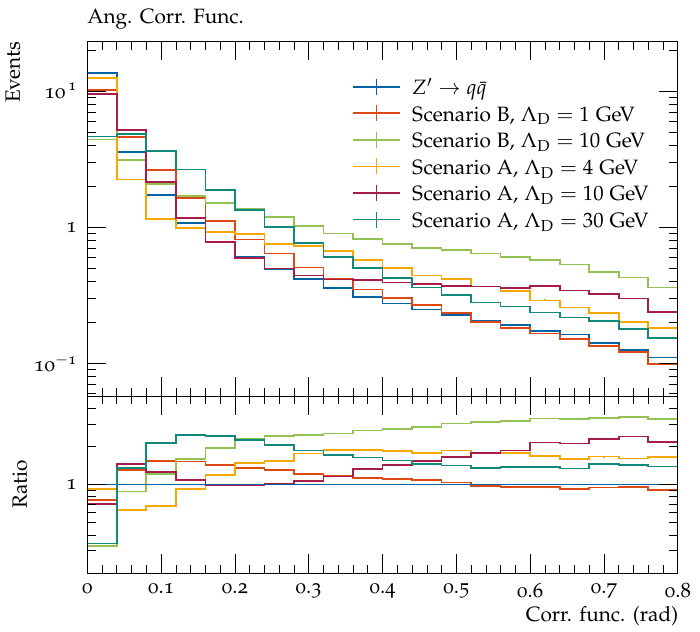}
    \caption{The correlation functions for all visible particles within each jet for the considered dark shower benchmarks and pair production of SM quarks by the $\zp$ mediator. All processes are normalised to unity. }
    \label{fig:corr_in_jets}
\end{figure}

One can also probe smaller angular separations by looking at the correlation functions within jets (figure \ref{fig:corr_in_jets}). For scenario A there is again a double peak structure visible in the ratio to QCD jets, with the first angular peak corresponding to the distance between the quarks from the dark hadron decays (in the first bin for \lamd=4 GeV, and the second to fourth bins for \lamd=10 GeV), and the second, broader peak coming from the distance between different dark hadrons. Note the second peak for the $\lamd=30$ GeV benchmark lies outside of the jet radius. For scenario B this two peak structure is not visible, which is likely due to a combination of two factors: firstly the decaying $\rhod$ mesons are higher in mass relative to $\lamd$, which controls the scale of the shower, so the two peaks are closer together, and secondly the majority of dark hadrons which decay are off-diagonal ones, which undergo three body decays, so the angular scale between the $q\bar{q}$ pair is not fixed, smearing the first peak. However a single clear angular scale, dependent on $\lamd$, is clearly visible.

\begin{figure}[!ht]
   \centering
    \includegraphics[width=0.45\textwidth]{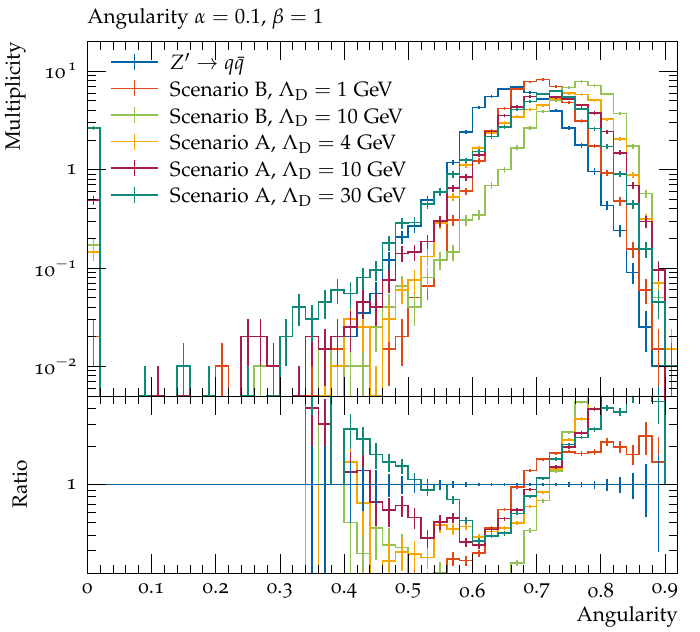}
    \includegraphics[width=0.45\textwidth]{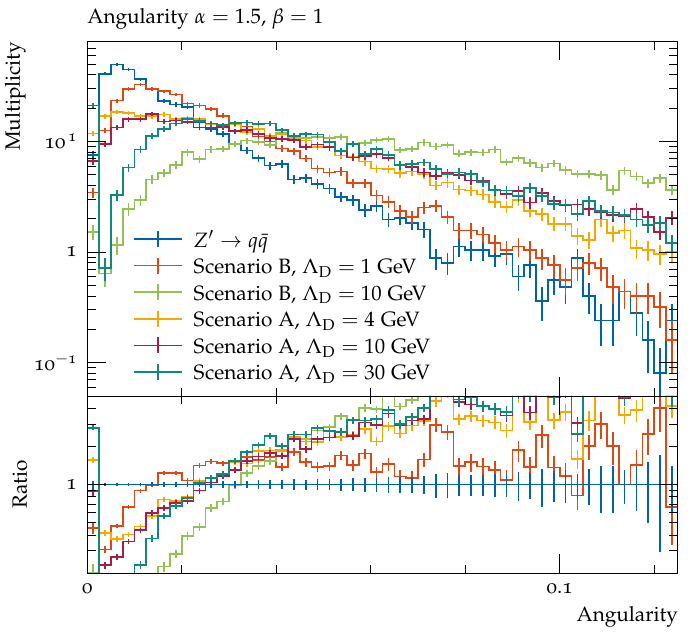}
     \caption{Angularities of visible jets with $\alpha=0.1, \beta=1$ (top), and $\alpha=1.5, \beta=1$ (bottom) for the considered dark shower benchmarks and pair production of SM quarks by the $\zp$ mediator. All processes are normalised to unity.}
    \label{fig:angularities}
\end{figure}

One can also see these angular structures using angularities, which have been proposed in \cite{angularities1,angularities2}. There are number of slightly different definitions for these variables, for instance those used in the previous references, but the ones used in this work are along the lines of \cite{Larkoski:2014uqa}, though using a standard jet axis:
\begin{eqnarray}
    \tau_{\alpha, \beta} &=& \sum_{\mathrm{jet}\in \mathrm{jets}}\sum_{i\in \mathrm{jet}}\left(2\sqrt{1-cos{\theta_{i}}} \right)^{\alpha}\left(\frac{E_{i}}{E_{Tot}}\right)^{\beta} \nonumber \\
    &\approx& \sum_{\mathrm{jet}\in \mathrm{jets}} \sum_{i\in \mathrm{jet}} \theta_{i}^{\alpha}\left(\frac{E_{i}}{E_{Tot}}\right)^{\beta}\ ,
\end{eqnarray}
where $\theta_{i}$ is the angle of the $i^{th}$ particle in a jet to the jet axis, $E_{i}$ is the energy of this particle, $E_{Tot}$ is the total energy in the event, and the final approximation holds for $\theta_{i} \ll 1$. These variables probe the structure of the jet at different angular and energy scales depending on the values of $\alpha$ and $\beta$ - small values of $\alpha$ probe particles close to the jet axis, which is sensitive to the distance between the decay products of individual light dark hadrons and larger values probe wide-angle emissions, which can correspond to both the angular scales of the decay products of heavier dark hadrons and the separation between dark hadrons for the lighter benchmarks. These observables are infrared safe for all $\alpha \geq 0$; here we consider $\alpha \in [0.1,2]$. Similarly small values of $\beta$ probe soft emissions, while higher values probe the harder particles in the jet, however these observables are infrared safe only for $\beta=1$, so we will consider this value here. 
As can be seen in figure \ref{fig:angularities}, the dark shower scenarios tend to higher values of angularities, as one would expect since the visible particles are more widely separated. This is most dramatic for the scenario B $\lamd=10$ GeV benchmark which has a large fraction of emissions at wide angles in the jets, however one can change the relative sensitivity by changing the angular power, with $\alpha=0.1$ improving the sensitivity to the lower \lamd{} benchmarks where the angular separation between the particles is on average smaller, and higher values of $\alpha$, such as 1.5, being particularly sensitive to the high \lamd{} benchmarks where there is a lot of wide-angle radiation.

\begin{figure}[!ht]
    \centering
    \includegraphics[width=0.45\textwidth]{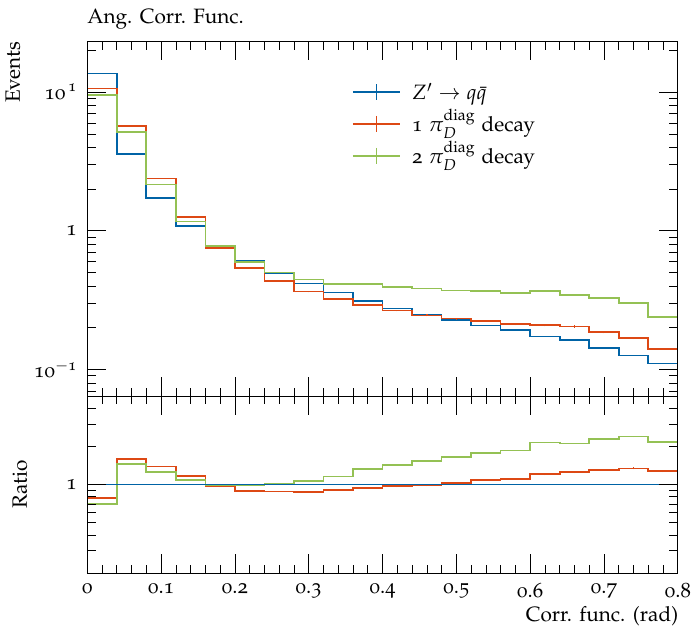}
    \caption{The correlation functions for all visible particles within each jet (weighted by the energy of the particles) for the scenario A \lamd=10 GeV benchmark in the cases that both or only one of the diagonal dark pions decay to SM particles, and pair production of SM quarks by the $\zp$ mediator. All processes are normalised to unity.}
    \label{fig:decay_nflav_corr}
\end{figure}

\begin{figure}[!ht]
    \centering
    \includegraphics[width=0.45\textwidth]{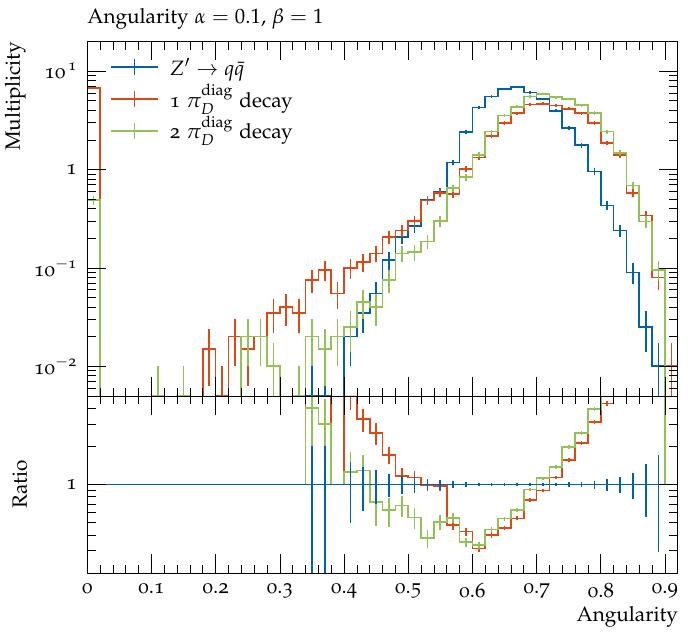}
    \includegraphics[width=0.45\textwidth]{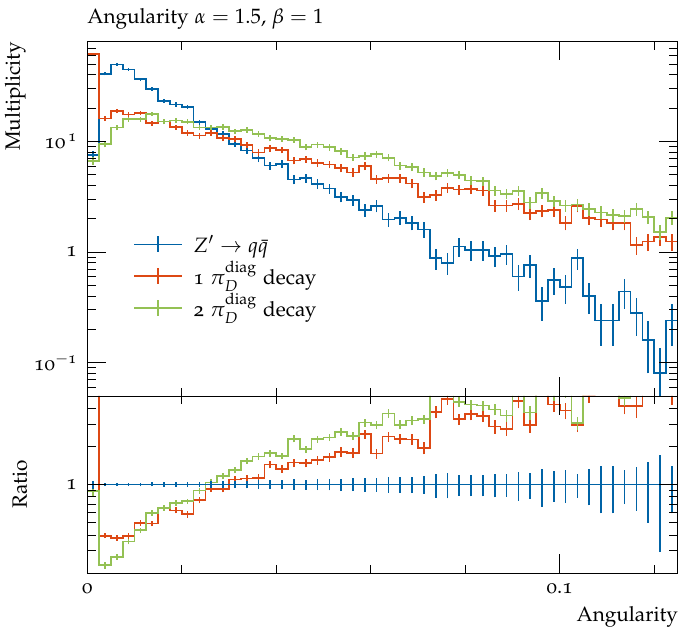}
     \caption{Angularities of visible jets with $\alpha=0.1, \beta=1$ (top), and $\alpha=1.5, \beta=1$ (bottom) for the scenario A \lamd=10 GeV benchmark in the cases that both or only one of the diagonal dark pions decay to SM particles, and pair production of SM quarks by the $\zp$ mediator. All processes are normalised to unity.}
    \label{fig:decay_nflav_angs}
\end{figure}

Finally we briefly investigated the impact of varying the number of diagonal dark pions which decay for the scenario A \lamd=10 GeV benchmark on the visible final state distributions. For this benchmark there are two diagonal dark pions, and so far we have considered the case that both decay to the SM. When only one is allowed to decay, there are fewer particles in the final state, which for the correlation functions (figure \ref{fig:decay_nflav_corr}) means the second peak, due to the angular distance between different dark hadrons is reduced, but the first peak, due to the distance between the decay products of a single hadron, is more prominent. Similarly for the angularities (figure \ref{fig:decay_nflav_angs}) little change is observed for $\alpha=0.1$ (except for an increase in events with very low angularities due to almost or completely invisible events), which is sensitive to the angular distance between decay products of a single dark hadron, while for $\alpha=1.5$ the distribution decreases more rapidly due to there being fewer dark hadrons decaying visibly to populate these wide angles.

\section{Impact of Parton Shower and Hadronisation Parameters}
\label{sec:had_param_var}

As discussed in section \ref{sec:psparams}, the parton shower and cluster hadronisation model have a number of parameters which must be set based on intuition from the Standard Model, however since these are not physical parameters the exact best values are often unclear. There are ongoing efforts to update the cluster hadronisation model to reduce the dependence on the parton shower cutoff and introduce a more physical model for cluster evolution \cite{Platzer:2022jny,Hoang:2024zwl}; these features would be particularly useful for dark shower predictions in the future, since one cannot tune the parameters to data, however we do not explore this in the context of dark showers for this study. In this section we instead investigate the effect of variations of the current parameters on the variables discussed in section \ref{sec:phenomenology}. For this study we focused only on the $\lamd=10$, scenario B benchmark, as the hadronisation parameters are particularly relevant for this case since they affect the fraction of \rhod{} mesons produced and hence the visible decay fraction.

\begin{figure}[!ht]
    \centering
  \begin{subfigure}[b]{.24\textwidth}
    \centering
    \includegraphics[width=1\textwidth]{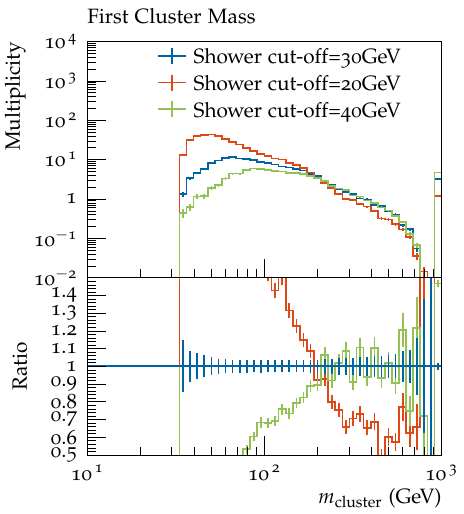}
    \caption{Initial cluster masses}
    \label{fig:first_clustermass_shower_CO_var}
\end{subfigure}\hfill
  \begin{subfigure}[b]{.24\textwidth}
    \centering
    \includegraphics[width=1\textwidth]{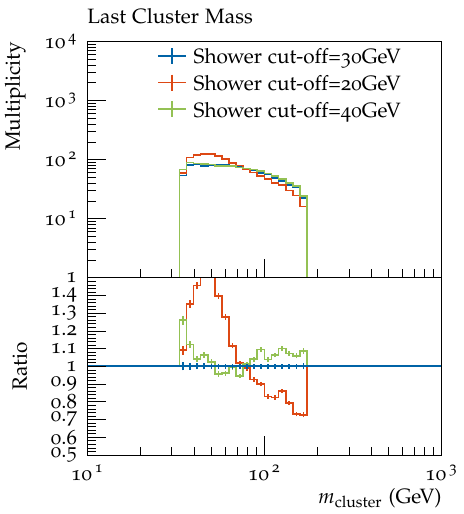}
    \caption{Final cluster masses}
    \label{fig:last_clustermass_shower_CO_var}
\end{subfigure}\hfill \\
\begin{subfigure}[b]{.24\textwidth}
    \centering
    \includegraphics[width=1\textwidth]{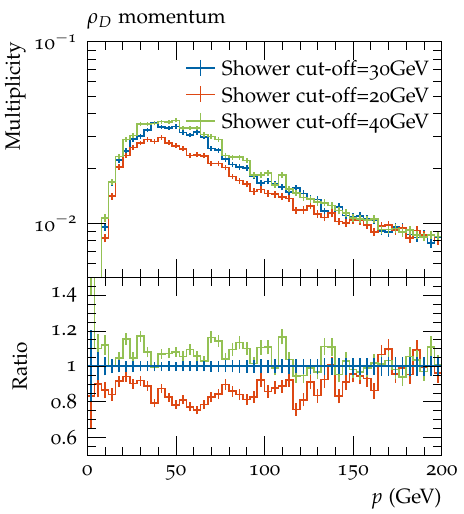}
    \caption{\rhod{} meson momentum}
    \label{fig:Rhod_p_shower_CO_var}
\end{subfigure} \hfill
\begin{subfigure}[b]{.24\textwidth}
    \centering
    \includegraphics[width=1\textwidth]{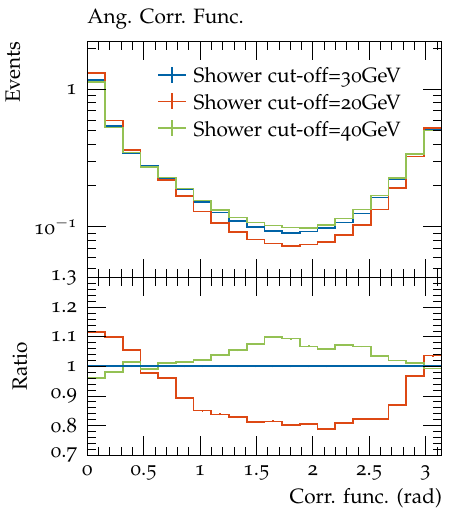}
    \caption{Correlation across event}
    \label{fig:CorrFunc_shower_CO_var}
\end{subfigure}
     \caption{Initial and final cluster masses, \rhod{} meson momentum and angular correlation across event for the $\lamd=10$, scenario B benchmark with the default settings (red) and shower cut-off variations.}
\end{figure}

The first parameter we investigated was the shower \pT{} cut-off, which we varied up and down by 10 GeV from the nominal value of 30 GeV. This is smaller than the corresponding range for which this parameter is tuned in the SM, where this parameter is varied and the hadronisation parameters tuned for each variation, however it is large enough to demonstrate the impact of this variable. Varying this parameter up cuts off many soft emissions, resulting in heavier clusters at the start of hadronisation, while varying it down results in more and lighter clusters (figure \ref{fig:first_clustermass_shower_CO_var}). After cluster fissioning the down variation still has significantly lighter clusters, though for the up variation the picture is less clear since while there are more of the heavy clusters close to the fission scale, the fissioning of the heavier clusters results in more of the lightest possible clusters as well (figure \ref{fig:last_clustermass_shower_CO_var}). Since decays to \rhod{}, and hence to visible particles, are supressed for cluster masses below $2\mrho$ and imposible for cluster masses less than $\mrho+\mpi$, the 20 GeV shower cut-off results in a notably lower fraction \rhod{} mesons of 50.5\%, compared to the benchmark value of 30 GeV (55.7\%) and the up variation of 40 GeV (55.5\%). The momentum distribution of the produced \rhod{} mesons also differs, with the down variation resulting in significantly softer \rhod{} mesons, while the up variation makes them slightly harder (figure \ref{fig:Rhod_p_shower_CO_var}). These two factors have opposite impacts on the final visible decay fraction: for the 20 GeV cut-off fewer \rhod{} mesons are produced but since they have higher momentum the final visible decay fraction is only slight lower, at 33.9\% (compared to 34.6\% for the benchmark), whereas for the 40 GeV cut-off the fact the rhos are softer means the visible decay fraction is also slightly lower, at 33.8\%. Looking at the visible distributions, the clearest impact is on the energy correlation function, where the up variation leads to a slight increase of particles at wide angles from each other, since there is more cluster fissioning, which tends to produce an isotropic distribution of particles, whereas the down variation results in lower initial cluster masses, less fissioning and hence more collimated distributions (figure \ref{fig:CorrFunc_shower_CO_var}).

\begin{figure}[!ht]
    \centering
  \begin{subfigure}[b]{.24\textwidth}
    \centering
    \includegraphics[width=1\textwidth]{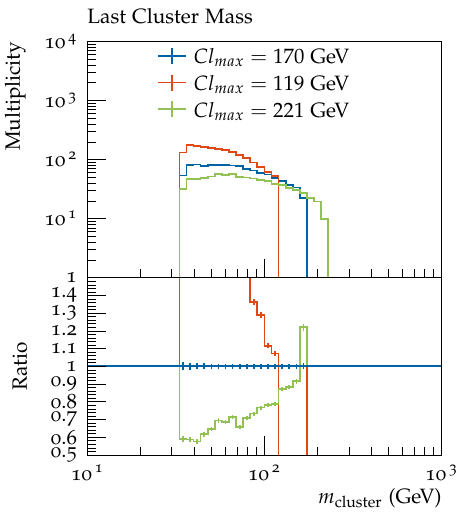}
    \caption{Final cluster masses}
    \label{fig:Clustermass_Clmax_var}
\end{subfigure}\hfill
\begin{subfigure}[b]{.24\textwidth}
    \centering
    \includegraphics[width=1\textwidth]{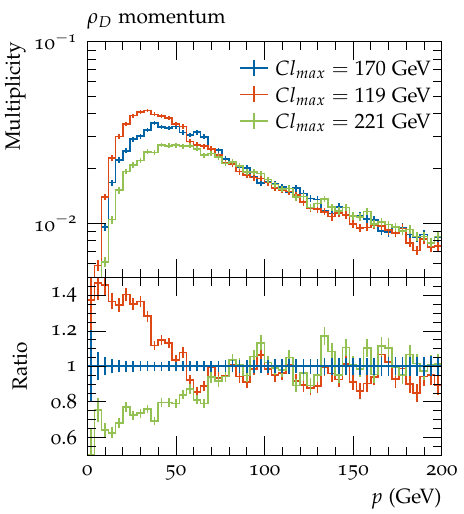}
    \caption{\rhod{} meson momentum}
    \label{fig:Rhod_p_Clmax_var}
\end{subfigure} \hfill \\
\begin{subfigure}[b]{.24\textwidth}
    \centering
    \includegraphics[width=1\textwidth]{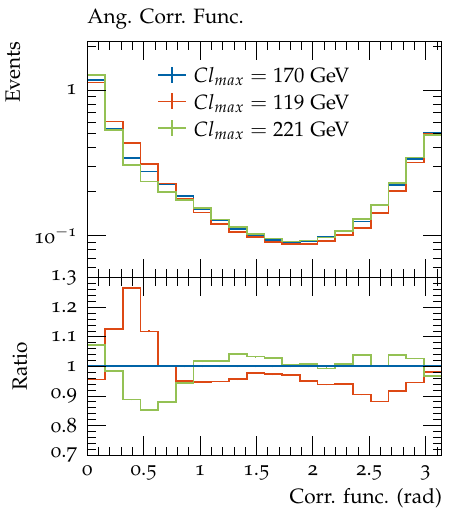}
    \caption{Correlation across event}
    \label{fig:CorrFunc_Clmax_var}
\end{subfigure}
     \caption{Final cluster masses, \rhod{} meson momentum and angular correlation across event for the $\lamd=10$, scenario B benchmark with the default settings (red) and $Cl_{\mathrm{max}}$ variations.}
\end{figure}

The next parameter considered was $Cl_{\mathrm{max}}$, which we varied by 30\%, based on the typical uncertainties when tuning this value to the Standard Model. Varying this parameter down results in the energy of the event being split across more light clusters after fissioning, while varying it up results in fewer, heavier clusters (figure \ref{fig:Clustermass_Clmax_var}). This significantly impacts the fraction of \rhod{} mesons, with a clear impact on the visible decay fraction - the default value of $Cl_{\mathrm{max}}=170$~GeV gives a \rhod{} fraction of 55.7\% and a visible decay fraction of 34.6\%, while the down variation gives 44.4\% and 29.3\% and the up variation 60.5\% and 37.0\% (though it should be noted that there are more \rhod{} mesons for the $Cl_{\mathrm{max}}$ down variation and fewer for the up variation due to the difference in the total number of clusters). Furthermore, the different cluster masses have a significant impact on the momenta of the \rhod{} mesons, with lower cluster masses resulting in decay products with less kinetic energy, while higher cluster masses result in higher average momentum (figure \ref{fig:Rhod_p_Clmax_var}). As a result of this, the angular separation of the decay products of individual \rhod{} mesons will tend to be higher for the $Cl_{\mathrm{max}}$ down variation and higher for the up variation, due to the differing boost of the system, which can be seen in the angular correlation functions (figure \ref{fig:CorrFunc_Clmax_var}) as a peak-dip structure in the ratio.

\begin{figure}[!ht]
    \centering
  \begin{subfigure}[b]{.24\textwidth}
    \centering
    \includegraphics[width=1\textwidth]{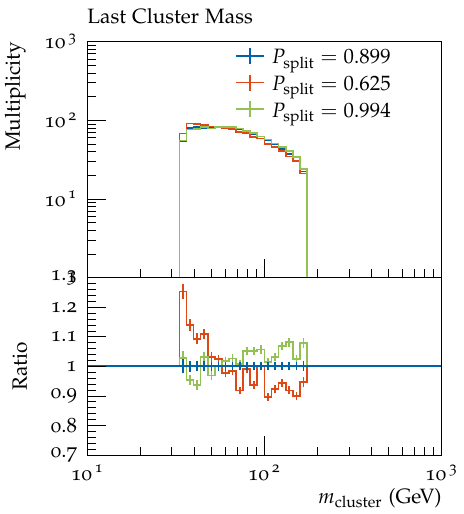}
    \caption{Final cluster masses}
    \label{fig:Clustermass_Psplit_var}
\end{subfigure}\hfill
\begin{subfigure}[b]{.24\textwidth}
    \centering
    \includegraphics[width=1\textwidth]{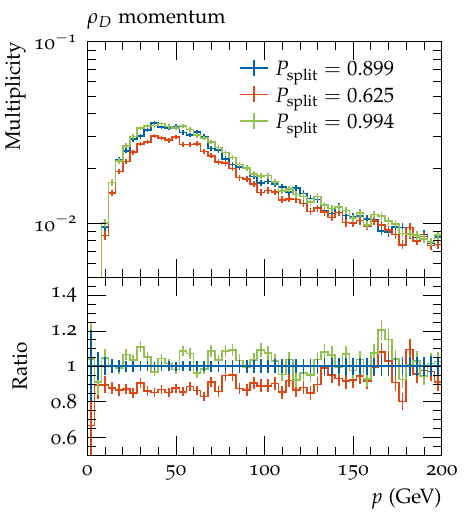}
    \caption{\rhod{} meson momentum}
    \label{fig:Rhod_p_Psplit_var}
\end{subfigure} \hfill \\
\begin{subfigure}[b]{.24\textwidth}
    \centering
    \includegraphics[width=1\textwidth]{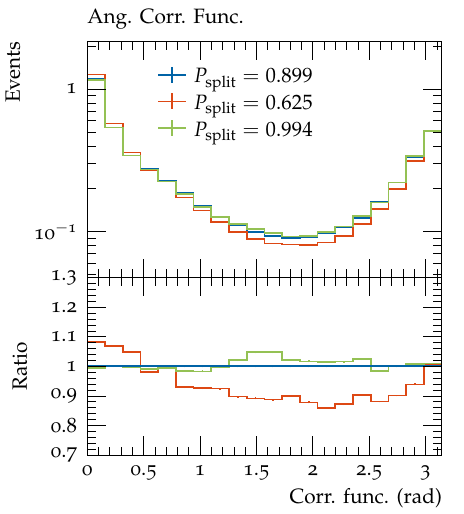}
    \caption{Correlation across event}
    \label{fig:CorrFunc_Psplit_var}
\end{subfigure}
     \caption{Final cluster masses, \rhod{} meson momentum and angular correlation across event for the $\lamd=10$, scenario B benchmark with the default settings (red) and \Psplit variations.}
\end{figure}

The next parameter investigated was \Psplit. We varied this parameter from its default value of 0.899 to the values for the c and b quarks, 0.994 and 0.625. Since the exact dynamics of dark cluster fissioning are unknown, the full range of sensible \Psplit{} values is hard to determine, but the range for different SM quarks seems a reasonable starting point. Varying this parameter down makes the spectrum of cluster masses (figure \ref{fig:Clustermass_Psplit_var}) after cluster fissioning slightly lighter, as more energy is converted to momentum of the clusters, while varying it up results in slightly more massive clusters. As a result the \rhod{} and hence visible energy fractions are lower for $\Psplit=0.625$ (54.3\% and 34.0\%), and higher for $\Psplit=0.994$ (56.4\% and 35.1\%). For the up variation the momentum of the \rhod{} meson shows little change, as the lower momentum of the clusters cancels the higher momentum given to the decay products of these clusters, but for the down variation the higher momentum of the clusters dominates, so the \rhod{} mesons are harder. This additional boost tends to give events a more back-to-back topology, as can be seen from the higher value of the angular correlation functions at small angles and lower value in the intermediate range (figure \ref{fig:CorrFunc_Psplit_var}). 

\begin{figure}[!ht]
    \centering
\begin{subfigure}[b]{.24\textwidth}
    \centering
    \includegraphics[width=1\textwidth]{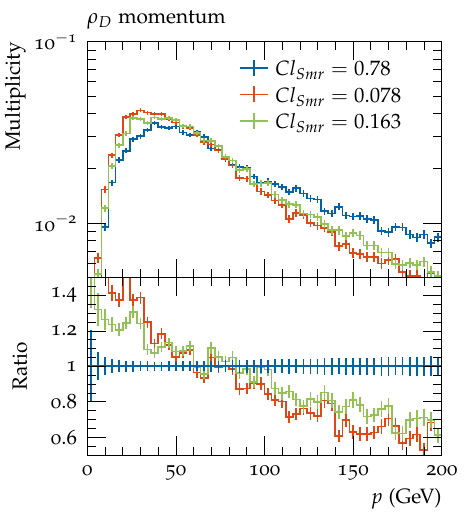}
    \caption{\rhod{} meson momentum}
    \label{fig:Rhod_p_ClSmr_var}
\end{subfigure} \hfill
\begin{subfigure}[b]{.24\textwidth}
    \centering
    \includegraphics[width=1\textwidth]{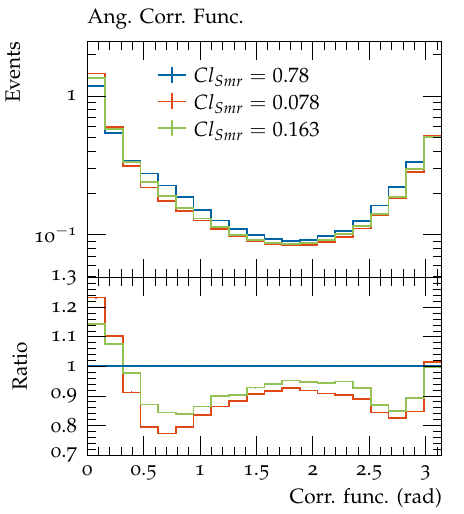}
    \caption{Correlation across event}
    \label{fig:CorrFunc_ClSmr_var}
\end{subfigure}
     \caption{\rhod{} meson momentum and angular correlation across event for the $\lamd=10$, scenario B benchmark with the default settings (red) and $Cl_{Smr}$ variations.}
\end{figure}

The final parameter which was varied was $Cl_{Smr}$, which we varied down from 0.78 to 0.163 and 0.078, which are the values for c and b quarks, respectively. Of these the c quark value of 0.163 is probably already quite low to be a reasonable parameter choice, since the dark quarks for this benchmark are much lighter relative to \lamd{} than even the c quark relative to $\Lambda_{QCD}$, but the b quark value is also included for illustrative purposes. Varying this parameter down reduces the momentum of the dark hadrons (figure \ref{fig:Rhod_p_ClSmr_var}).  The angular correlation functions across the event (figure \ref{fig:CorrFunc_ClSmr_var}) also show an interesting shape - this is likely due to the dark rho mesons in the two jets being more aligned, so the decay products within a single jet have less angular separation, while the angular separation to the objects in the other jet is higher. 

\section{Summary and Outlook}
\label{conclusion}

We have implemented the generation of dark shower events in the Herwig generator, which will be included in the upcoming Herwig 7.4 release. This is the first dark shower event generator to use an angular ordered shower, and the first to use a cluster model for dark hadronisation. Wherever possible, the existing Herwig code structure has been extended so that future developments can also be directly used for dark shower predictions. So far it is possible to generate models with a confining force controlled by an $SU(N_{C})$ symmetry for arbitrary $N_{C}>2$ and $2 \leq \nf \leq 9$ dark quarks, however extensions to further scenarios are foreseen in future releases.

In this paper we also analysed a set of physically motivated benchmark points as outlined in \cite{Albouy:2022cin}, including a sensible set of parameters for generating these models with Herwig. We then outlined variables which can be used to internally validate the generation, for instance the masses of the clusters in the hadronisation model, and variables which can be used to help discriminate against SM QCD backgrounds. In particular angular correlation functions show promise for studying these events, and may help for understanding the phenomenology and devising search strategies.

Finally, we investigated the impact of varying the parton shower and hadronisation parameters from the SM-inspired values we initially selected. It is clear that in the current hadronisation model varying these parameters, especially the shower cut-off and $Cl_{\mathrm{max}}$ variables can have a significant impact on the angular distributions and dark decay fractions, while the \Psplit{} and $Cl_{Smr}$ parameters have smaller, but still notable effects. Ongoing developments for the cluster hadronisation model should give a better theoretical grounding, which should make it easier to select well-motivated values for the hadronisation parameters, and provide a method for estimating the uncertainty from the hadronisation model.
 
\section*{Acknowledgements}

SK is supported by the FWF research group funding FG1 and FWF project
number P 36947-N. MRM is supported by the UK Science and Technology
Facilities Council (grant numbers ST/P001246/1). DS acknowledges
support from DESY (Hamburg, Germany), a member of the Helmholtz Association
HGF. SP would like to thank Wouter Waalewijn and Massimiliano Procura for fruitful
discussions on angularities. We thank the other Herwig team members
for discussions and technical advice, and Andreas Papaefstathiou and
Andrzej Siódmok in particular for discussions at an early stage of
this project. We thank Matt Strassler for careful reading and comments on the draft. 

\bibliographystyle{spphys}
\bibliography{refs} 

\begin{thebibliography}{10}
\providecommand{\url}[1]{{#1}}
\providecommand{\urlprefix}{URL }
\expandafter\ifx\csname urlstyle\endcsname\relax
  \providecommand{\doi}[1]{DOI \discretionary{}{}{}#1}\else
  \providecommand{\doi}{DOI \discretionary{}{}{}\begingroup
  \urlstyle{rm}\Url}\fi

\bibitem{Strassler:2006im}
M.J. Strassler, K.M. Zurek, Phys. Lett. \textbf{B651}, 374 (2007).
\newblock \doi{10.1016/j.physletb.2007.06.055}

\bibitem{Han:2007ae}
T.~Han, Z.~Si, K.M. Zurek, M.J. Strassler, JHEP \textbf{07}, 008 (2008).
\newblock \doi{10.1088/1126-6708/2008/07/008}

\bibitem{Kulkarni:2022bvh}
S.~Kulkarni, A.~Maas, S.~Mee, M.~Nikolic, J.~Pradler, F.~Zierler, SciPost Phys.
  \textbf{14}, 044 (2023).
\newblock \doi{10.21468/SciPostPhys.14.3.044}

\bibitem{Beauchesne:2019ato}
H.~Beauchesne, G.~Grilli~di Cortona, JHEP \textbf{02}, 196 (2020).
\newblock \doi{10.1007/JHEP02(2020)196}

\bibitem{Pomper:2024otb}
J.~Pomper, S.~Kulkarni,   (2024)

\bibitem{Hochberg:2014kqa}
Y.~Hochberg, E.~Kuflik, H.~Murayama, T.~Volansky, J.G. Wacker, Phys. Rev. Lett.
  \textbf{115}(2), 021301 (2015).
\newblock \doi{10.1103/PhysRevLett.115.021301}

\bibitem{Kamada:2022zwb}
A.~Kamada, S.~Kobayashi, T.~Kuwahara, JHEP \textbf{02}, 217 (2023).
\newblock \doi{10.1007/JHEP02(2023)217}

\bibitem{Batz:2023zef}
A.~Batz, T.~Cohen, D.~Curtin, C.~Gemmell, G.D. Kribs, JHEP \textbf{04}, 070
  (2024).
\newblock \doi{10.1007/JHEP04(2024)070}

\bibitem{Soni:2016gzf}
A.~Soni, Y.~Zhang, Phys. Rev. D \textbf{93}(11), 115025 (2016).
\newblock \doi{10.1103/PhysRevD.93.115025}

\bibitem{Curtin:2022oec}
D.~Curtin, C.~Gemmell, JHEP \textbf{09}, 010 (2023).
\newblock \doi{10.1007/JHEP09(2023)010}

\bibitem{Forestell:2016qhc}
L.~Forestell, D.E. Morrissey, K.~Sigurdson, Phys. Rev. D \textbf{95}(1), 015032
  (2017).
\newblock \doi{10.1103/PhysRevD.95.015032}

\bibitem{Soni:2017nlm}
A.~Soni, H.~Xiao, Y.~Zhang, Phys. Rev. D \textbf{96}(8), 083514 (2017).
\newblock \doi{10.1103/PhysRevD.96.083514}

\bibitem{Carenza:2022pjd}
P.~Carenza, R.~Pasechnik, G.~Salinas, Z.W. Wang, Phys. Rev. Lett.
  \textbf{129}(26), 261302 (2022).
\newblock \doi{10.1103/PhysRevLett.129.261302}

\bibitem{Cohen:2015toa}
T.~Cohen, M.~Lisanti, H.K. Lou, Phys. Rev. Lett. \textbf{115}(17), 171804
  (2015).
\newblock \doi{10.1103/PhysRevLett.115.171804}

\bibitem{Cohen:2017pzm}
T.~Cohen, M.~Lisanti, H.K. Lou, S.~Mishra-Sharma, JHEP \textbf{11}, 196 (2017).
\newblock \doi{10.1007/JHEP11(2017)196}

\bibitem{Beauchesne:2017yhh}
H.~Beauchesne, E.~Bertuzzo, G.~Grilli Di~Cortona, Z.~Tabrizi, JHEP \textbf{08},
  030 (2018).
\newblock \doi{10.1007/JHEP08(2018)030}

\bibitem{ArkaniHamed:2008qp}
N.~Arkani-Hamed, N.~Weiner, JHEP \textbf{12}, 104 (2008).
\newblock \doi{10.1088/1126-6708/2008/12/104}

\bibitem{Baumgart:2009tn}
M.~Baumgart, C.~Cheung, J.T. Ruderman, L.T. Wang, I.~Yavin, JHEP \textbf{04},
  014 (2009).
\newblock \doi{10.1088/1126-6708/2009/04/014}

\bibitem{Chan:2011aa}
Y.F. Chan, M.~Low, D.E. Morrissey, A.P. Spray, JHEP \textbf{05}, 155 (2012).
\newblock \doi{10.1007/JHEP05(2012)155}

\bibitem{Buschmann:2015awa}
M.~Buschmann, J.~Kopp, J.~Liu, P.A.N. Machado, JHEP \textbf{07}, 045 (2015).
\newblock \doi{10.1007/JHEP07(2015)045}

\bibitem{Schwaller:2015gea}
P.~Schwaller, D.~Stolarski, A.~Weiler, JHEP \textbf{05}, 059 (2015).
\newblock \doi{10.1007/JHEP05(2015)059}

\bibitem{Renner:2018fhh}
S.~Renner, P.~Schwaller, JHEP \textbf{08}, 052 (2018).
\newblock \doi{10.1007/JHEP08(2018)052}

\bibitem{Polchinski:2002jw}
J.~Polchinski, M.J. Strassler, JHEP \textbf{05}, 012 (2003).
\newblock \doi{10.1088/1126-6708/2003/05/012}

\bibitem{Hatta:2008tx}
Y.~Hatta, E.~Iancu, A.H. Mueller, JHEP \textbf{05}, 037 (2008).
\newblock \doi{10.1088/1126-6708/2008/05/037}

\bibitem{Knapen:2016hky}
S.~Knapen, S.~Pagan~Griso, M.~Papucci, D.J. Robinson, JHEP \textbf{08}, 076
  (2017).
\newblock \doi{10.1007/JHEP08(2017)076}

\bibitem{Harnik:2008ax}
R.~Harnik, T.~Wizansky, Phys. Rev. \textbf{D80}, 075015 (2009).
\newblock \doi{10.1103/PhysRevD.80.075015}

\bibitem{CMS:2021dzg}
A.~Tumasyan, et~al., JHEP \textbf{06}, 156 (2022).
\newblock \doi{10.1007/JHEP06(2022)156}

\bibitem{ATLAS:2023swa}
G.~Aad, et~al.,   (2023)

\bibitem{Albouy:2022cin}
G.~Albouy, et~al., Eur. Phys. J. C \textbf{82}(12), 1132 (2022).
\newblock \doi{10.1140/epjc/s10052-022-11048-8}

\bibitem{Kribs:2016cew}
G.D. Kribs, E.T. Neil, Int. J. Mod. Phys. A \textbf{31}(22), 1643004 (2016).
\newblock \doi{10.1142/S0217751X16430041}

\bibitem{Cacciapaglia:2020kgq}
G.~Cacciapaglia, C.~Pica, F.~Sannino, Phys. Rept. \textbf{877}, 1 (2020).
\newblock \doi{10.1016/j.physrep.2020.07.002}

\bibitem{Carloni:2010tw}
L.~Carloni, T.~Sj{\"{o}}strand, JHEP \textbf{09}, 105 (2010).
\newblock \doi{10.1007/JHEP09(2010)105}

\bibitem{Carloni:2011kk}
L.~Carloni, J.~Rathsman, T.~Sj{\"{o}}strand, JHEP \textbf{04}, 091 (2011).
\newblock \doi{10.1007/JHEP04(2011)091}

\bibitem{Bierlich:2022pfr}
C.~Bierlich, et~al., SciPost Phys. Codeb. \textbf{2022}, 8 (2022).
\newblock \doi{10.21468/SciPostPhysCodeb.8}

\bibitem{Platzer:2022jny}
S.~Pl\"atzer, JHEP \textbf{07}, 126 (2023).
\newblock \doi{10.1007/JHEP07(2023)126}

\bibitem{Hoang:2024zwl}
A.H. Hoang, O.L. Jin, S.~Pl\"atzer, D.~Samitz,   (2024)

\bibitem{Bahr:2008pv}
M.~Bahr, et~al., Eur. Phys. J. C \textbf{58}, 639 (2008).
\newblock \doi{10.1140/epjc/s10052-008-0798-9}

\bibitem{Bellm:2015jjp}
J.~Bellm, et~al., Eur. Phys. J. C \textbf{76}(4), 196 (2016).
\newblock \doi{10.1140/epjc/s10052-016-4018-8}

\bibitem{Bewick:2023tfi}
G.~Bewick, et~al.,   (2023)

\bibitem{Gieseke:2003rz}
S.~Gieseke, P.~Stephens, B.~Webber, JHEP \textbf{12}, 045 (2003).
\newblock \doi{10.1088/1126-6708/2003/12/045}

\bibitem{Platzer:2009jq}
S.~Platzer, S.~Gieseke, JHEP \textbf{01}, 024 (2011).
\newblock \doi{10.1007/JHEP01(2011)024}

\bibitem{Platzer:2011bc}
S.~Platzer, S.~Gieseke, Eur. Phys. J. C \textbf{72}, 2187 (2012).
\newblock \doi{10.1140/epjc/s10052-012-2187-7}

\bibitem{Masouminia:2021kne}
M.R. Masouminia, P.~Richardson, JHEP \textbf{04}, 112 (2022).
\newblock \doi{10.1007/JHEP04(2022)112}

\bibitem{Darvishi:2021het}
N.~Darvishi, M.R. Masouminia, Nucl. Phys. B \textbf{985}, 116025 (2022).
\newblock \doi{10.1016/j.nuclphysb.2022.116025}

\bibitem{Lee:2023hef}
J.B. Lee, M.R. Masouminia, M.H. Seymour, U.k. Yang, JHEP \textbf{08}, 064
  (2024).
\newblock \doi{10.1007/JHEP08(2024)064}

\bibitem{Richardson:2018pvo}
P.~Richardson, S.~Webster, Eur. Phys. J. C \textbf{80}(2), 83 (2020).
\newblock \doi{10.1140/epjc/s10052-019-7429-5}

\bibitem{Cormier:2018tog}
K.~Cormier, S.~Pl\"atzer, C.~Reuschle, P.~Richardson, S.~Webster, Eur. Phys. J.
  C \textbf{79}(11), 915 (2019).
\newblock \doi{10.1140/epjc/s10052-019-7370-7}

\bibitem{Hoang:2018zrp}
A.H. Hoang, S.~Pl\"atzer, D.~Samitz, JHEP \textbf{10}, 200 (2018).
\newblock \doi{10.1007/JHEP10(2018)200}

\bibitem{Bewick:2019rbu}
G.~Bewick, S.~Ferrario~Ravasio, P.~Richardson, M.H. Seymour, JHEP \textbf{04},
  019 (2020).
\newblock \doi{10.1007/JHEP04(2020)019}

\bibitem{Bewick:2021nhc}
G.~Bewick, S.~Ferrario~Ravasio, P.~Richardson, M.H. Seymour, JHEP \textbf{01},
  026 (2022).
\newblock \doi{10.1007/JHEP01(2022)026}

\bibitem{Webber:1983if}
B.R. Webber, Nucl. Phys. B \textbf{238}, 492 (1984).
\newblock \doi{10.1016/0550-3213(84)90333-X}

\bibitem{Herwigpp}
M.~Bahr, et~al., Eur. Phys. J. C \textbf{58}, 639 (2008).
\newblock \doi{10.1140/epjc/s10052-008-0798-9}

\bibitem{Gieseke:2017clv}
S.~Gieseke, P.~Kirchgae\ss{}er, S.~Pl\"atzer, Eur. Phys. J. C \textbf{78}(2),
  99 (2018).
\newblock \doi{10.1140/epjc/s10052-018-5585-7}

\bibitem{Fischer:2006ub}
C.S. Fischer, J. Phys. G \textbf{32}, R253 (2006).
\newblock \doi{10.1088/0954-3899/32/8/R02}

\bibitem{Matchbox}
S.~Platzer, S.~Gieseke, Eur. Phys. J. C \textbf{72}, 2187 (2012).
\newblock \doi{10.1140/epjc/s10052-012-2187-7}

\bibitem{Herwig7}
J.~Bellm, et~al., Eur. Phys. J. C \textbf{76}(4), 196 (2016).
\newblock \doi{10.1140/epjc/s10052-016-4018-8}

\bibitem{Rivet}
A.~Buckley, J.~Butterworth, D.~Grellscheid, H.~Hoeth, L.~Lonnblad, J.~Monk,
  H.~Schulz, F.~Siegert, Comput. Phys. Commun. \textbf{184}, 2803 (2013).
\newblock \doi{10.1016/j.cpc.2013.05.021}

\bibitem{Cohen:2020afv}
T.~Cohen, J.~Doss, M.~Freytsis, JHEP \textbf{09}, 118 (2020).
\newblock \doi{10.1007/JHEP09(2020)118}

\bibitem{angularities1}
L.G. Almeida, S.J. Lee, G.~Perez, G.F. Sterman, I.~Sung, J.~Virzi, Phys. Rev. D
  \textbf{79}, 074017 (2009).
\newblock \doi{10.1103/PhysRevD.79.074017}

\bibitem{angularities2}
C.F. Berger, T.~Kucs, G.F. Sterman, Phys. Rev. D \textbf{68}, 014012 (2003).
\newblock \doi{10.1103/PhysRevD.68.014012}

\bibitem{Larkoski:2014uqa}
A.J. Larkoski, D.~Neill, J.~Thaler, JHEP \textbf{04}, 017 (2014).
\newblock \doi{10.1007/JHEP04(2014)017}

\end{thebibliography}

\end{document}